# Evaluating Smart Home Device Users’ Responses to their (Un)Confirmed Privacy Expectations

TANIA KHATUN, Louisiana State University, USA

MAHDIEH SHEIKH REZAEI, Louisiana State University, USA

DANNY YUXING HUANG, New York University, USA

ODED NOV, New York University, USA

REZA GHAIUMY ANARAKY, Louisiana State University, USA

Users of smart home devices are often unaware of how their devices handle personal data. We examine how revealing these data practices influences users’ trust, satisfaction, and coping behaviors, including decisions to block device communications. Using Expectation-Confirmation Theory, we conducted two complementary studies to balance ecological validity with experimental control. An in-situ field study ($N = 35$) used network monitoring to reveal actual device traffic, and an online experiment ($N = 132$) presented simulated reports with manipulated levels of advertising-related communications. Across both studies, when data practices aligned with users’ expectations, satisfaction increased, strengthening intentions to continue using the device. Defensive responses, however, followed different pathways: satisfaction predicted willingness to block in the in-situ field study, whereas collection concerns were the primary predictor of blocking in the experiment. Together, these findings show how transparency reshapes attitudes and behaviors among existing smart-home users, underscoring the role of expectation confirmation in real-world, continued-use contexts. CCS Concepts: • Human-centered computing → Empirical studies in ubiquitous and mobile computing; • Security and privacy → Usability in security and privacy.

## 1 Introduction

Smart home and Internet of Things (IoT) devices are changing people’s interactions with their living environments [32, 82]. These devices enable users to monitor and control various systems in their homes, from adjusting room lights and temperature via voice commands to managing home security systems [4, 35, 83]. These benefits, however, come at the potential cost of users’ privacy; studies show that smart home users are concerned that smart devices may compromise their privacy [16]. Therefore, for an effective, responsible, and user-sensitive deployment of smart home systems, it is crucial to understand the expectations and privacy concerns of smart home users and develop ways to address them.

One reason for users’ privacy concerns is the lack of transparency about the data practices of smart home devices (i.e., how they handle user data) [41, 97]. Following Janic et al. [51], transparency refers to providing users with accurate and comprehensible insight into how their data are collected, stored, processed, and disclosed. Under this definition, transparency may be insufficient when relevant information about smart-home devices’ data practices is omitted, incomplete, or presented in a way that is difficult for users to understand. Knowledge of smart-home devices’ data practices is a driver of user trust [41, 79], whereas limited transparency about data practices can erode users’ trust [51, 55].

Despite the importance of transparency, companies often fail to communicate the data practices of their smart home products to consumers. Sometimes, they fail to present the information in the first place [49, 73]; for example, Google

Authors’ Contact Information: Tania Khatun, tkhatu1@lsu.edu, Louisiana State University, Baton Rouge, Louisiana, USA; Mahdieh Sheikh Rezaei, msheik5@lsu.edu, Louisiana State University, Baton Rouge, Louisiana, USA; Danny Yuxing Huang, dhuang@nyu.edu, New York University, New York, New

York, USA; Oded Nov, onov@nyu.edu, New York University, New York, New York, USA; Reza Ghaiumy Anaraky, [reza-email], Louisiana State University, Baton Rouge, Louisiana, USA.

1

failed to disclose the presence of a microphone chip in the Nest Secure hub to the consumers [12]. At other times, companies fail to present the information in an understandable manner [49]. For example, privacy policies that include information on how user data is collected and used are difficult to read and comprehend [67, 74], and prior studies have documented limited engagement with, and awareness of, privacy policy content among their participants [5, 8]. Therefore, prior scholarly literature has argued that manufacturers and developers often fall short of this transparency by either omitting relevant information about data practices or presenting it in ways that are difficult for users to understand [41, 49, 53, 60, 80].

Although insufficient transparency is an industry-wide problem that requires manufacturer-side design changes and policy-level interventions, understanding how users respond when device data practices are made visible is a necessary empirical step toward informing those interventions. To surface these data practices and provide users with greater transparency, scholars proposed centralized solutions to users for managing their privacy and security [27, 47, 50, 65]. Such solutions involve using tools that monitor users' network traffic to reveal the data practices of their smart home devices. However, the consequences of achieving transparency with such tools and for devices already in use have not been well-studied. While smart home device users may trust their devices to some degree to act in their best interests, it is unclear if achieving transparency and learning about the data practices of one's smart home devices influences one's attitudes (e.g., trust and satisfaction) and behaviors (e.g., intention to continue using devices and willingness to block future device communications) about their devices. In this work, we attempt to address this gap and answer the following research question:

RQ: How does revealing the data practices of smart home devices influence users' attitudes and behaviors towards these devices?

To address our research question, our study is guided by the Expectation-Confirmation Theory (ECT) [11]. ECT posits that users have expectations of a technology prior to adopting it, and if they find that the technology meets their initial expectations (i.e., confirmation of the expectations), they are satisfied with the technology and are more likely to continue using it [11]. On the other hand, if the technology fails to meet users' expectations, users' satisfaction and their behavioral intention to continue using it are reduced. In this study, we explore how induced transparency can confirm users' trust expectations in their smart home devices. Additionally, we investigate whether trust confirmation (or lack thereof) predicts users' satisfaction with their smart home devices, their willingness to block device communications, and their intention to continue using these devices.

We conducted two studies to address this question. In the first study, we notified smart home device users about the data practices of their devices using *IoT Inspector*, a network monitoring software tool that generates a report on the network traffic communications of smart home devices present in one's network [47]. This report includes data about the frequency at which the smart home devices communicate data, the data destination domain's name (e.g., google.com), and the destination geographic location (e.g., USA). In the study, we present a screenshot of this tool in Figure 2. Additionally, the tool asks users to help researchers detect anomalies by selecting the communications between each device and domain that they wish to block (bottom-right side of Figure 2). IoT Inspector is an open-source software and can be installed on Windows, Mac, and Linux devices. We conducted this study with participants who signed up to use the IoT Inspector upon release. However, since this initial participant pool was limited, we also conducted an experimental study with a larger sample from Prolific. In the second study, smart device users viewed a simulated report on the data practices of one of their smart home devices and answered questions about their expectations and whether they were confirmed.

Results from both our in-situ field study and controlled experiment consistently support the Expectation-Confirmation Theory: confirming trust expectations enhances satisfaction, which in turn strengthens the intention to continue using the device. However, our findings highlight a critical divergence from the ECT: while satisfaction mediated the decision to continue usage in both studies, the predictors of the intended active protective response differed: blocking a device's future communications was associated with satisfaction in the field study, but driven primarily by data collection concerns in the controlled experiment. This distinction suggests that while satisfaction promotes retention, specific risk perceptions are what trigger protective controls in some populations. This paper makes several contributions to the literature:

- Prior smart-home privacy research has examined users' mental models and concerns through interviews [99] and elicited privacy judgments about hypothetical information flows [7]. Closest to our work is Aretha [85], a qualitative technology probe that exposed three households to real traffic from their own devices. Our work complements these studies through a two-study design that combines in-situ exposure to reports generated from participants' actual device communications (study 1) with a controlled experiment using simulated reports (study 2). Study 1 provides ecologically grounded evidence from smart-home users in their own homes, whereas Study 2 provides experimental control over the level of advertising-related communication. Together, the studies allow us to apply ECT to examine post-adoption trust confirmation, satisfaction, continued-use intention, and intended privacy-protective responses across complementary in-situ and experimental contexts.
- Prior transparency research has often focused on simplifying privacy information for users before adoption or purchase, for example, through nutrition labels [38, 56] or comics [5]. Consistent with Janic et al.'s [51] definition of transparency, these approaches primarily provide users with accurate and comprehensible information about data practices at the point of purchase or adoption. Our work extends this literature to post-adoption contexts by examining how users respond when information about device communications becomes visible after they have already incorporated a smart-home device into their lives. This focus allows us to examine outcomes beyond purchase decisions, including trust confirmation, continued-use intention, and willingness to block future communications. This approach informs the design of future transparency mechanisms and manufacturer- and policy-level decisions about when and how to communicate data practices.
- Our work builds on ECT by exploring a coping response to unfavorable device communications that are not captured by continued-use intentions alone. Lowering intentions to continue usage is a passive coping behavior and is predicted in the ECT [11]. However, users whose privacy expectations are violated may employ other strategies to address this violation. Studying such alternatives is crucial as it provides insights into designing technologies that continue to benefit users without compromising their privacy or necessitating reduced usage. Specifically, we examine users' stated willingness to engage in one active, control-oriented response: selecting unfavorable communication instances that they would be willing to block and prevent from recurring in the future. This is an initial exploratory examination of how ECT may help explain a more active coping response.

## 2 Literature Review

In the following sections, we first review the broader literature on the data practices of smart home devices, followed by a summary of work on transparency and control within smart home devices. Finally, we review the literature that studied privacy technologies through the lens of ECT.

### 2.1 Sharing Data with Smart Home Devices

To ensure users make informed choices about their smart home device data sharing, we need to understand what data collection approaches users are comfortable with and uncomfortable with. Therefore, scholars studied parameters that can predict data disclosure behaviors [6, 7, 7, 17, 63]. For example, Lee and Kobsa [63] asked participants to imagine IoT devices collecting information in various scenarios. These scenarios include information about the data collection's location (e.g., public place), the type of information collected (e.g., video), the recipient of information (e.g., government), the reason for sharing (e.g., safety), and the frequency of disclosure (once vs. continuously). They found that the extent to which participants are comfortable with a given data collection scenario depends on these variables, with the comfort level often declining as the scenarios become more intrusive (e.g., continuous disclosure rather than a one-time disclosure).

Another factor predicting disclosure tendencies is whether or not the collecting parties share users' data with others [70]. Users perceive using smart devices as riskier if such devices share data with third parties [24, 100], whether the third parties are Internet Service Providers (ISP), governments, or advertisement companies [100]. Sharing data with third parties can make users' data prone to misuse [71]. Furthermore, users sometimes believe they may not benefit from sharing data with third parties. For example, Malkin et al. [71] found that users uniformly distrust ISPs with their smart home data since they believed that ISPs could not deliver any additional benefits to them. Another common case of disclosing user data to third parties is data for targeted advertisements. Although receiving an ad about a product that may be beneficial and relevant to the user can be appealing (i.e., the targeted advertising), it comes at the cost of sharing user data with the ad providers. Therefore, many users are worried about sharing data with ad domains [21], yet many remain unaware that their data is shared with external parties via third-party apps [69]. The situation is more alarming if users do not expect their interaction data to be used for advertising purposes. For example, Iqbal et al. [49] found that
Amazon's smart speaker uses user interaction data to learn about users' interests for the purpose of targeted advertising. IoT users' attitudes towards their devices and their sharing behaviors can potentially change if they become aware of such unexpected data practices. In particular, prior work highlights a complex interplay between who receives the data and the specific purposes for which the device is used, both of which strongly influence users' disclosure decisions [30].

Smart home devices can collect or reveal many types of information, including audio, video, usage patterns, device interaction logs, routines, location-related information, and physical presence [100]. Although these data types can all have privacy and security implications, users are often especially concerned about audio and video data. For example, the majority of users are very uncomfortable sharing identifiable data or video, while being more comfortable sharing data that communicates their physical presence [78]. Audio data is also perceived as sensitive and can be collected by smart speakers. However, whether these devices listen to users' conversations at home remains a subject of public debate [1, 2]. This concern may stem from a lack of transparency and uncertainty about whether users can trust smart speakers' data practices. Indeed, in addition to the contextual variables discussed above, the lack of transparency and control over data is a significant privacy issue for users of smart devices [62]. Below, we review the literature on transparency and control.

### 2.2 Transparency, an Ever-lasting Privacy Issue

Transparency is a fundamental principle that especially should be respected in privacy-sensitive technologies [81]. However, developers do not always put effort into promoting transparency. More than half of the apps operating on Alexa's ecosystem do not have privacy policies on their install page, and most of the ones with a privacy policy are too generic and apply to several products of the same developer [49]. Additionally, transparency is hard to achieve [52]; it requires users to invest time and effort into reading device documentation and policies [75]. These documents are often too complicated for the average user to understand [67], motivating scholars to seek ways to promote transparency.

There have been several attempts to promote transparency by simplifying user privacy information. Kelly et al. [56] were the first to propose the use of privacy nutrition labels that summarize important privacy information. Ghaiumy Anaraky et al. [5] used comics to make the privacy information more engaging to users. Gideon et al. [36] designed a "privacy-enhanced search engine" that annotates results with the privacy policy information from the websites. In the context of smart devices, Emami-Naeini et al. [23] interviewed privacy experts about what information should be available on the labels of smart devices. As a result, they designed nutrition labels for smart devices that highlight important information such as sensor type (e.g., audio, temperature), the purpose of data collection (e.g., providing functionality, research), and whether the data is shared with third parties. Despite such efforts to promote transparency, research shows that users' perceptions about the data practices of their smart devices do not always align with reality. Al-Ameen et al. [3] conducted a lab study and found a mismatch between smart device users' perceptions and the devices' actual data practices. For example, around 30% of users believed that their smart homes do not collect their personal information (e.g., name, gender, and date of birth), while the device collected such information. It is noteworthy that this data collection is not always more intrusive than users may expect. In the same study, Al-Ameen et al. [3] found that 58.33% users thought their IoT device stores their data forever, while in practice, the devices discarded the data after a period of time.

Transparency and trust are intertwined, as users are more likely to trust a transparent system that communicates its data practices to them [19]. Research outlines three dimensions for trust: competence (entity's expertise in the subject matter), benevolence (entity's tendencies to act in consumers' best interests), and integrity (entity being dependable) [76]. In scenarios where users decide to disclose their personal information to providers, benevolence and integrity are more influential than competence [76]. Specifically, benevolence is identified as a key variable in consumer behavior research [9, 28], marketing, [31, 86], and privacy literature [39, 61]. Benevolence is the strongest predictor of user satisfaction among other dimensions of trust [96] and predicts usage intentions [94], two other constructs of our interest in this study. Guo et al. studied how transparent and non-transparent privacy policies predict perceived benevolence and found that users find entities more benevolent if they have transparent privacy policies [39]. In this work, we study how transparency, as the realization of data practices of one's smart home devices, influences one's perceived benevolence of their smart home devices.

### 2.3 Users' Control Over Smart Sensors' Pervasive Data Collection

One of the primary sources of user concern for smart devices is the lack of control over how these devices stream users' data to the cloud [20]. Research shows that users employ various ways to address this concern [99]. Zeng et al.[99] interviewed fifteen smart home device users and identified two major privacy management strategies. The first strategy was a non-technical risk mitigation strategy where users changed their behavior towards their smart home devices (e.g., avoiding saying sensitive things around smart speakers). The second strategy was technical risk mitigation, where users blocked some traffic or used a separate Wi-Fi for some devices. Our work explores both types of strategies. We study changes in intentions to use smart home devices in the future as a non-technical mitigation strategy. Additionally, we delve deeper into technical strategies and study the usage of data control mechanisms. Since exercising control separately on each device is effortful and confusing, scholars proposed centralized privacy choice platforms to help users manage their data more efficiently [27, 100]. For example, Zheng et al. [100] conducted an interview study with smart home users and asked about potential control mechanisms. Their participants appreciated a feature where they could see commands recorded by Amazon Echo and delete some of them, which they were uncomfortable with. The authors suggest developing control applications that enable users to specify which data their devices can collect and upload to the internet and to view and delete the already-collected data. However, the efficacy of such centralized control systems and the extent to which smart home users would use these platforms to change the privacy behaviors of their smart devices in real life is unclear. In this work, we explore whether users are willing to mark the data transactions that they want to block in the future.

Taken together, prior work has examined smart-home privacy through several complementary approaches. Zeng et al. [99] used interviews to elicit users' mental models, concerns, and mitigation strategies, including network segmentation and changes to device settings. Apthorpe et al. [7], drawing on Contextual Integrity theory, elicited privacy judgments about hypothetical information flows through a crowdsourced survey. In these studies, participants reasoned about device data practices rather than viewing network activity generated by their own devices. Seymour et al. [85] used Aretha, a technology probe that showed participants actual network traffic from their own devices during a six-week deployment with three households. Aretha provides valuable qualitative insights and design implications, whereas our work combines an in-situ field study with a controlled experiment and applies ECT to examine post-adoption trust confirmation, satisfaction, continued-use intention, and intended privacy-protective responses.

# 3 Theoretical Framework and Hypotheses Development

## 3.1 ECT in the Privacy Literature

ECT explains users' post-adoption intentions to continue using a technology [11]. In Bhattacherjee's informationsystems continuance model, users evaluate whether their experience with a technology confirms their prior expectations; confirmation shapes satisfaction, which in turn predicts continuance intention. The model distinguishes users' intention to continue using a system after adoption from the initial adoption decision and focuses on post-acceptance variables, including confirmation and satisfaction, to explain continued use. In this paper, we apply ECT to smart-home privacy by narrowing the relevant expectation to a privacy-specific construct: users' subjective belief that their smart-home devices and associated data-handling ecosystems act in their best interests when handling personal data. This psychological use of "expectation" differs from the legal concept of consumers' "reasonable expectations." For example, Kustosch et al. [59] examine what consumers may legitimately expect from IoT manufacturers as a basis for consumer protection and manufacturer accountability, whereas our study examines the expectations users actually hold and how their confirmation or disconfirmation relates to subsequent responses. Our contribution is therefore behavioral rather than legal, and we do not evaluate manufacturers' legal obligations.

Prior privacy research has applied ECT to privacy-sensitive technologies [40, 92].Wang et al. [92] studied the purchase intention of health products on WeChat and showed that those who find their expectations confirmed are more likely to report a higher intention to purchase other products on the business system. However, they measured confirmation in general terms (e.g., "The product and service provided by this system were better than what I had expected.") and were not able to map confirmation to various aspects of the user's experience. Gupta et al. [40] showed that if mobile wallets confirm users' performance and efficacy expectations, users report higher satisfaction and intentions for future usage. Similarly, while they measured different types of expectations (performance expectations and effortfulness expectations), they measured the confirmation in broad terms (e.g., "The basic services provided by the M-wallet are better than what I expected"), making it impossible to relate confirmation to specific aspects of the user's experience.

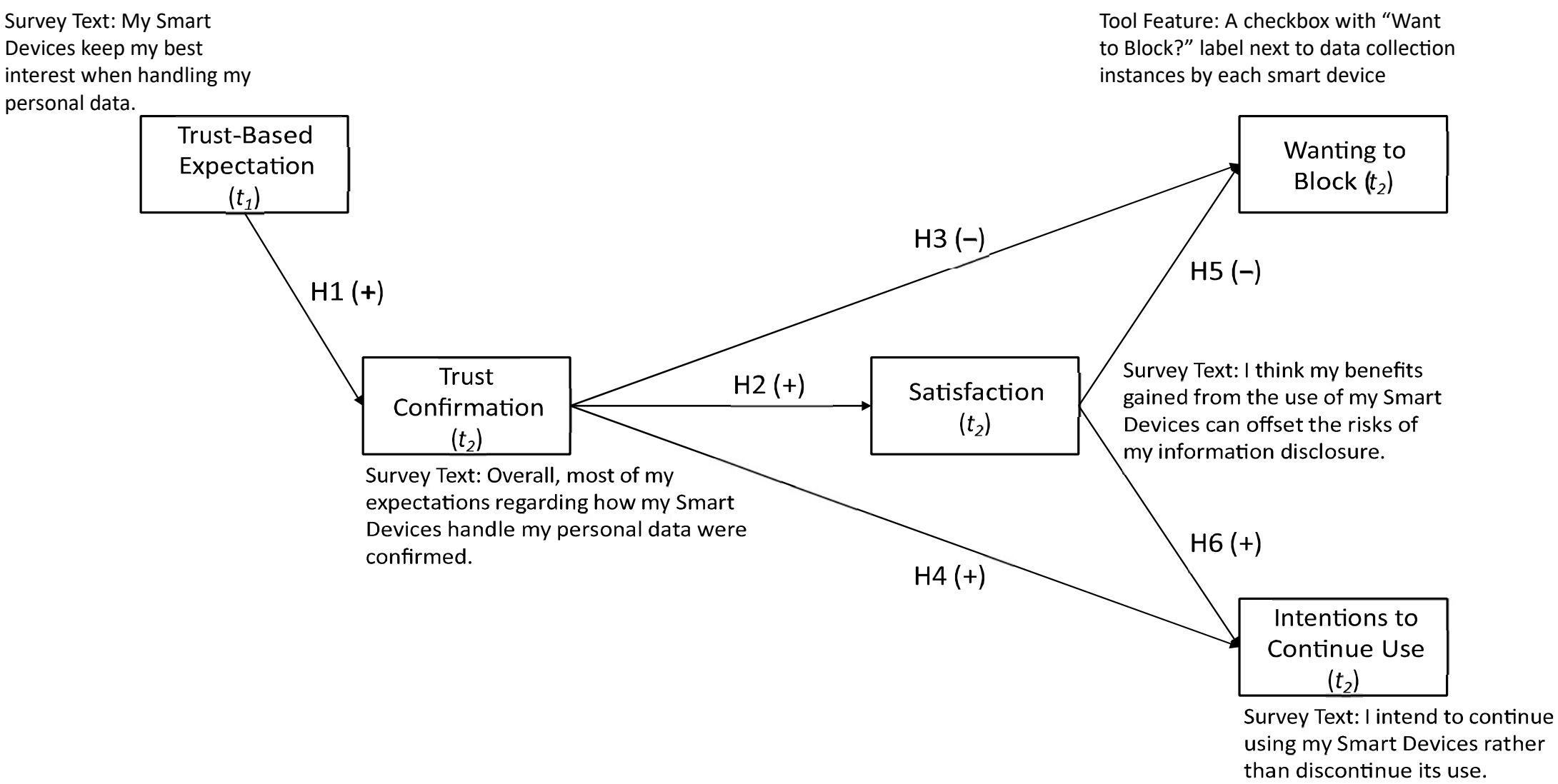

Fig. 1. Hypothesized model. Trust-based expectation was measured before participants viewed the report ($t_1$), whereas trust confirmation, satisfaction, continued-use intention, and willingness to block were measured afterward ($t_2$). The figure includes the Study 1 survey item or interface control associated with each construct and the expected direction of each hypothesized relationship. We have included the text of questions associated with each construct for enhanced readability.

In this work, we specifically focus on users' expectations of smart home devices keeping their best interests when handling their data, and navigating the consequences of confirming or not confirming such expectations.

## 3.2 Hypothesis Development

The privacy literature provides ample evidence suggesting that privacy policies are not readable by average users and that users are unaware of the data practices of the technologies they use [5, 67]. This leads to a gap between users' expectations about data practices of their technologies, which is based on their understanding (or misunderstanding) of the privacy policies, and the actual data practices of the technology they use [3]. This situation can arguably be more severe in the context of smart home devices because such devices operate pervasively in the users' living spaces without users' active attention. Therefore, users may forget that the devices operate in the background and communicate data to various parties. In addition, these device communications are not always easily noticed by users. For example, it may be easier to notice data disclosures on a social media platform (e.g., seeing the location shared upon publishing a post) than to notice data communicated by a smart thermostat. Therefore, we argue that users' expectations of data practices of their smart home devices are not aligned with how the devices actually handle users' data, and users' expectations of devices being benevolent and trustworthy are not confirmed after seeing the IoT Inspector's report. However, since posing a null hypothesis is unconventional, we hypothesize a positive association between users' trust-based expectations about their smart home devices and the confirmation of these expectations.
H1: Users who trust their smart home devices are more likely to find their trust confirmed after seeing the IoT Inspector's report.

However, the extent to which users' trust-based expectations are confirmed or disconfirmed would predict their consequent satisfaction with their smart home devices. A low level of confirmation suggests that smart home devices are poorly transparent in terms of privacy and do not keep up with users' expectations. Poor transparency is associated with low satisfaction [54]. On the contrary, users of highly transparent systems know about the underlying data practices of the system and report higher levels of satisfaction with the system [26]. In line with these findings, we argue that the extent to which users' initial expectations are confirmed would predict their satisfaction with their smart home devices: H2: The more users' trust in their smart home devices is confirmed, the more they will be satisfied with their devices.

When users' initial expectations are not confirmed, they may cope with the perceived violation of their privacy in various ways. First, they may limit their smart device use: social media research has shown that users whose privacy has been violated reduce their social media use [10]. Overall, users tend to limit their interactions with the devices that they perceive as unsafe [22]. Another possible response is to use privacy controls to limit undesirable data communications. Such controls may allow users to exert greater control over their data and better align device practices with their privacy preferences [18, 72]. As a narrow operationalization of a more active, control-oriented response, we examine users' stated willingness to use checkboxes to mark undesirable device communications for future blocking.
H3: The more users' trust in smart home devices is confirmed, the less they will want to block their devices' communications.
H4: The more users' trust in their smart home devices is confirmed, the more they want to continue using their smart home devices.

Satisfaction is consistently reported as a predictor of behavioral intentions [11]. For example, in the context of online banking, higher satisfaction is associated with a higher intention to use the online banking system in the future [89]. In line with this literature, we pose the following hypotheses:
H5: The more users are satisfied with their smart home devices, the less they want to block their devices' communications.
H6: The more users are satisfied with their smart devices, the more they want to continue using smart home devices.

Figure 1 summarizes our hypothesized model.

# 4 Study 1: The In-Situ Field Study

## 4.1 Methods

### 4.1.1 *IoT Inspector.*

We used the IoT Inspector, an open-source tool designed by researchers across various universities. This software enables users to monitor their smart home devices' network communications [47]. The software tracks network traffic, reporting details on destination domains and their geolocations, as well as the frequency of communications with third-party advertising and tracking services. Figure 2 shows a screenshot of IoT Inspector. Users can download this software for free and install it on Windows, Mac, and Linux operating systems. IoT Inspector has built-in mechanisms to protect the privacy of its participants, such as collecting only the headers of packets in aggregate and omitting what appears to be non-IoT devices. Interested readers can read the original papers [45, 47], its IRB protocols [44], and the source code [46] for more details.

### 4.1.2 *Procedure.*

IoT Inspector is a public domain tool. It has a public website that Google can crawl [48], and its downloadable binaries and source code are hosted on a public repository [46]. Additionally, multiple news outlets and social media platforms published stories about IoT Inspector [29, 58]. IoT Inspector's website has a "waiting list" that interested users can sign up for with their email addresses to be notified of updates and new releases. Therefore, IoT Inspector's users could learn about the software from any of the means above and come from anywhere in the world with various degrees of technical expertise. Because participants had voluntarily joined the IoT Inspector waiting list and installed the tool, they may have been more interested in IoT privacy and security or more technically motivated than the general public. However, we did not measure their motivations and therefore cannot infer that they had higher privacy concerns. To learn more about IoT Inspector's user base, please refer to Huang et al. [45, 47].

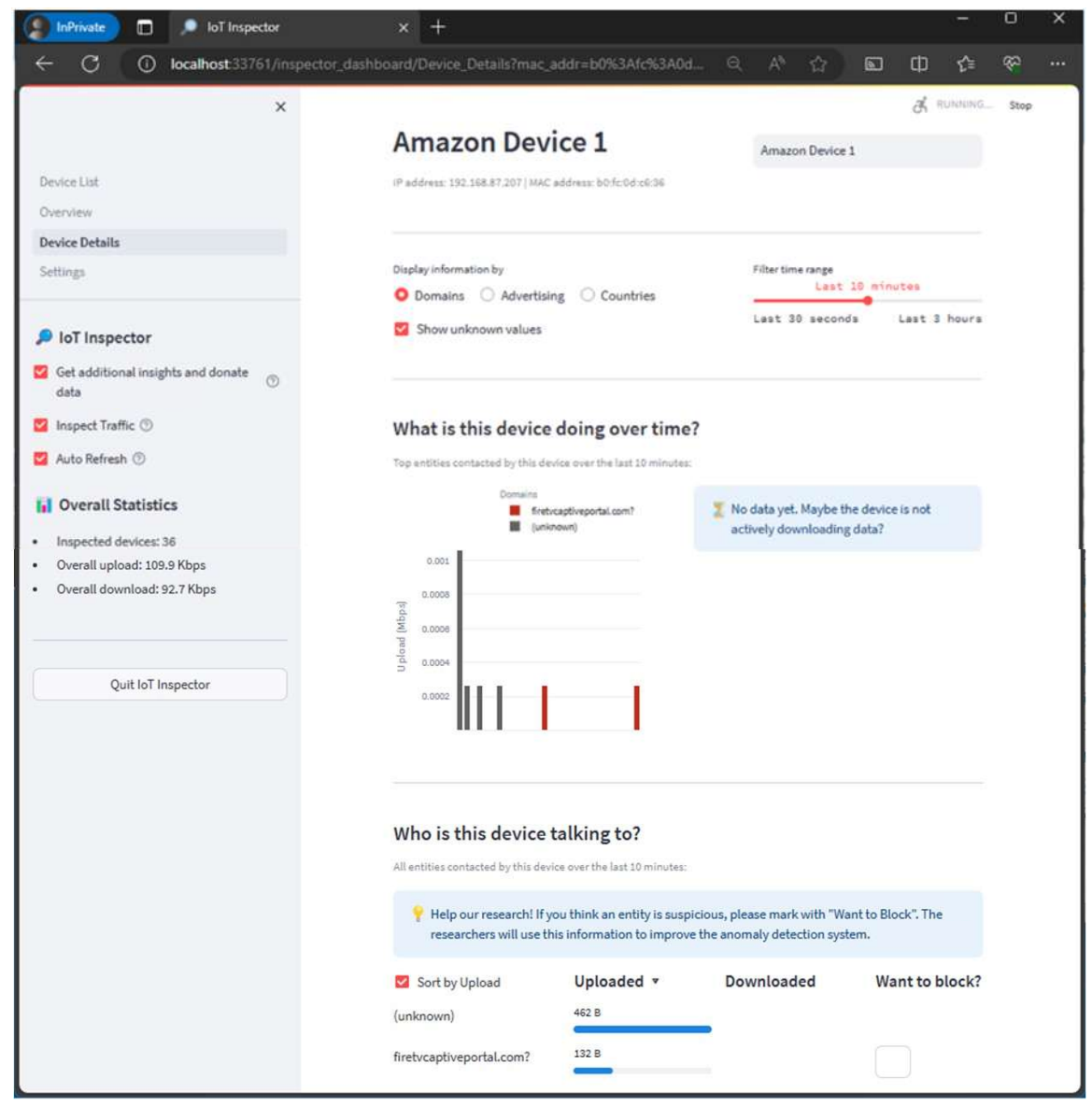

Fig. 2. A screenshot of the IoT Inspector. Users install the software and access the user interface through their browser.

With approval from our Institutional Review Board (IRB) and the IoT Inspector research team's IRB [44], we aimed to recruit participants from individuals who signed up on IoT Inspector's waiting list. The IoT Inspector team contacted everyone on the waiting list, letting them know that the IoT Inspector has become available for use. After learning about the tool's release and installing IoT Inspector, users saw an informed consent form asking them whether they wanted to participate in our study. The users who rejected the consent could still use the IoT Inspector without answering survey questions or donating network data. The IoT Inspector operated locally for these users without sending us any data.

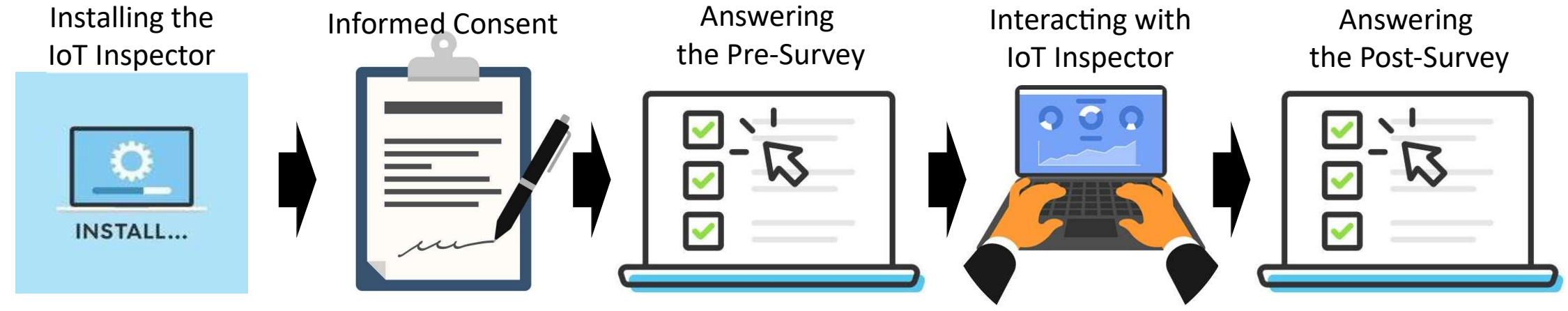

Fig. 3. Users installed the IoT Inspector and saw the informed consent. Those who did not consent could still use the tool without answering surveys or sharing data. Those who gave consent were immediately redirected to answer the pre-survey. Then, they could interact with the tool and view the reports. After five minutes, we prompted them to respond to the post-survey.

We only collected data from those who gave consent to answer survey questions and donate data (N = 35 participants finished the study). In the consent form, we communicated that data donation includes sharing the IoT Inspector's report, including information about their smart devices, the frequency at which these devices send out data, and the domains with which the devices communicate.

We administered two surveys; users answered the first survey before they saw the IoT Inspector's report about their smart home devices. Then, after users browsed and navigated the IoT Inspector's report for at least five minutes, we prompted them with a pop-up window and directed them to the second survey. If participants attempted to close the app before completing the post-survey, we prompted them with the same pop-up window and reminded them to answer the post-survey before leaving the app. Therefore, participants who dismissed the initial pop-up without completing the post-survey and later attempted to close the application could see the pop-up a second time. Because the study's analyses required paired pre- and post-survey responses, participants who completed only the pre-survey could not be included in the path model. The prompt, therefore, served as a reminder intended to reduce post-survey dropout in this autonomous field deployment; participants could still exit without completing the survey. Please see Figure 3 for a summary of our experimental protocol.

Our survey drew upon several constructs from the existing literature. We tried to keep the survey as short as possible since our study did not include monetary compensation, and longer questionnaires can lower response rates [84]. Retaining participants was especially important because our participant pool was limited to individuals who had voluntarily signed up to use IoT Inspector, and we could not supplement a low response rate with a new participant pool. Therefore, we selected one item from each construct of interest. We recognize that single-item measures do not allow for the assessment of internal reliability or discriminant validity and provide less robust operationalizations of rich psychological constructs, such as trust confirmation and satisfaction, than multi-item scales. As a pragmatic approach to minimizing participant burden in this uncompensated field deployment, we selected either the highest-loading item from the original validated instrument or, based on discussion among the authors, the item that most directly captured the construct's conceptual definition. Responses were recorded using a 7-point Likert scale ranging from "Strongly Disagree" to "Strongly Agree." Below, we present the pre- and post-interaction measures.

Pre Survey Instruments:

- Trust-based expectation: We measured users' expectations of smart devices following trustworthy data practices. We used the benevolence dimension of trust, which is concerned with whether an entity acts in the user's best interests or not [93]. We borrowed this item from previous work [93] and tailored it into the context of smart home privacy: "My Smart Devices keep my best interests when handling my personal data." Because the trust items explicitly asked participants about how their smart-home devices handle personal data, trust in this study should be understood as users' expectations that the device and its associated manufacturer and service ecosystem will collect, process, and share data in ways aligned with users' interests. It should not be interpreted as general trust in the device's functionality or narrowly as trust in its technical security.
- Privacy concerns: We included a single-item measure of global privacy concern as a baseline control variable, using an item from the Internet Users' Information Privacy Concerns (IUIPC) scale [70]: "All things considered, the Internet would cause serious privacy problems." We included this instrument to capture participants' general privacy concern as a person-level privacy orientation. Including this measure as a control is consistent with prior privacy research that uses general privacy concern as an individual-difference control when examining more context-specific privacy attitudes and behaviors [15]. Our goal was to control for participants' general privacy orientation so that the associations among trust-based expectations, trust confirmation, satisfaction, continued-use intention, and blocking behavior could be interpreted above and beyond this broader tendency.

Post Survey Instruments:

- Trust confirmation: We followed the literature's approach in measuring confirmation [11] and asked users if their trust expectations of their smart devices were confirmed: "Overall, most of my expectations regarding how my Smart Devices handle my personal data were confirmed." Trust-confirmation items were coded such that higher scores indicate stronger perceived confirmation of participants' prior trust-based expectations, whereas lower scores indicate weaker confirmation.
- Satisfaction: The economic theory of utility posits that individuals attempt to achieve satisfaction or maximum utility through their choices, given their resource limitations. Xu et al. [95] define perceived value as a comprehensive assessment of the choice object in the decision-making process. In our context, the "choice object" is the use of the smart-home device in light of the data practices made visible to the participant. The "comprehensive assessment" refers to users' overall judgment of whether the benefits of using the device are sufficient relative to the privacy risks associated with those data practices; it does not imply an exhaustive evaluation of every device feature or possible use. The item was adapted from Xu et al.'s [95] Perceived Value of Information Disclosure scale: "I think my benefits gained from the use of my Smart Devices can offset the risks of my information disclosure." Accordingly, this operationalization does not capture broad affective satisfaction with the device, as measured in some canonical ECT formulations [11]. Instead, it captures satisfaction in a privacy-focused context: users' evaluative judgment of whether the benefits of device use are sufficient relative to the privacy risks of disclosure.

- Continuous Intention: We followed the literature's approach, measuring continuous intention to use a technology [11] and tailored it to the context of smart home devices: "I intend to continue using my Smart Devices rather than discontinue their use."

*4.1.3 Participant Recruitment.* Overall, 152 users agreed to participate in the study and completed the pre-survey questionnaire. However, only 35 individuals also completed the post-survey questionnaire.[1] Therefore, our analyses include the 35 participants who completed both surveys. Participation rates are often lower in field studies with limited incentives (e.g., 11% in a field study with a $2 incentive [87]). This low completion rate should be interpreted in light of Study 1's uncompensated field-deployment design. Participants installed and used IoT Inspector on their own and completed the post-survey only after interacting with the tool. Although the available pre-survey measures did not indicate differences between completers and non-completers, the groups may still have differed in unmeasured ways, which may limit the generalizability of Study 1. We also did not collect demographic information for Study 1 and therefore cannot characterize the sample demographically. Study 2 partially addresses this limitation by collecting and reporting participants' demographic characteristics.

*4.1.4 Post-hoc Power Analysis:* We conducted a post-hoc power analysis for the individual equations predicting the two dependent variables. At an alpha level of 0.05 and with a sample of 35 participants, the estimated power was 0.85 for predicting intention to continue use and 0.97 for predicting willingness to block a device communication [88]. These estimates pertain to the individual equations predicting the two dependent variables and should not be interpreted as evidence that the full path model in Study 1 was adequately powered.

*4.1.5 Data Analysis.* We conducted a path model with our hypothesized effects [66]. Path models can inform us about the mechanisms and pathways through which an effect occurs. Additionally, they help us explore potential mediations and whether and how the relationship between two variables (e.g., confirmation and intention to use) is channeled through a third variable (e.g., satisfaction). In the case of mediation, we report indirect effects that show the strength of the indirect association. Furthermore, we conducted a saturated model to study if there are other significant nonhypothesized effects. Since we did not find any additional significant effects, we do not report the results of the saturated model. We did not explore any two-way interaction effects, as we did not have enough statistical power for more complicated analyses. Given the small sample size and the use of single-item measures in Study 1, we interpret the path model as exploratory evidence of associations among the measured variables rather than as a confirmatory test of latent constructs. Study 2, with its larger sample and multi-item measures, provides a stronger statistical and measurement test.

### 4.2 Results

Participants' initial trust-based expectations of their smart home devices varied, such that six users strongly disagreed that their devices kept their best interests in mind when handling their personal data, indicating low trust-based expectations regarding how the smart-device ecosystem would handle their data. Seven users disagreed with the statement, two users somewhat disagreed, nine users neither agreed nor disagreed, one user somewhat agreed, nine agreed, and one strongly agreed. After viewing the IoT Inspector report, participants reported their trust confirmation , with two users disagreeing about the presence of an alignment between their initial expectations and what they found in IoT Inspector's report about the actual data practices of their smart devices; one user somewhat disagreed, four users were neutral, five users somewhat agreed, twenty users agreed, and three users strongly agreed. Participants' privacy concern, measured on a scale from 1 to 7, had a mean of 5.61 ($SD$ = 1.34), suggesting relatively high concern in our sample. However, the privacy-concern measure was not significantly associated with any other variable in our model. Therefore, we removed it from the analyses (please see the discussion section for a more elaborate discussion about privacy concerns).

[1] We compared the 35 participants who completed both surveys with the 117 participants who completed only the pre-survey on the variables measured at pre-survey. We found no significant differences between the groups on these measured variables; however, this analysis cannot rule out differences in unmeasured characteristics.

*4.2.1 Hypotheses Testing.* We conducted a path model with all our hypotheses (see Figure 4). The results of the path model of study 1 are presented in Table 1, where statistically significant hypotheses are presented in bold. This model shows an acceptable fit ($\chi^2(10) = 95.435$, $p < .001$, CFI = 0.996, TLI = 0.988, SRMR = 0.038, RMSEA = 0.054, p = .386, CI : [0.000, 0.295]) [34]. The results suggest that users who trusted their smart devices more, were also more likely to report a higher confirmation of their trusting beliefs, such that by one standard deviation increase in users' initial trust at t1, their reported trust confirmation increased by 0.885 standard deviations ($p < .001$, H1 confirmed). Additionally, we found that the more users' trust is confirmed, the more they feel satisfied with their smart devices. By one standard deviation increase in confirming having truthful devices, the satisfaction increases by 0.409 standard deviations ($p = 0.002$, H2 confirmed). However, the results suggest that trust confirmation does not lead to significant changes in users' willingness to block their devices' communications ($B = -0.024$, $p = .466$, H3 rejected) or intentions to continue using smart home devices in the future ($B = 0.226$, $p = .067$, H4 rejected). Finally, those who reported higher satisfaction with their smart devices were less likely to want to block certain communications from their smart home devices ($B = -1.263$, $p < .001$, H5 confirmed) and more likely to want to continue using their devices in the future ($B = 0.395$, $p < .008$, H6 confirmed).

Table 1. The results of the path model of study 1. Significant hypotheses are presented in bold format.

| Variable Names | Standardized coefficients | Standard error | p-value |
|---|---|---|---|
| DV: Trust Confirmation (t2) | | | |
| **H1: Trust Expectations (t1)** | **0.885** | **0.078** | **<0.001** |
| DV: Satisfaction (t2) | | | |
| **H2: Trust Confirmation (t2)** | **0.409** | **0.140** | **0.002** |
| DV: Intentions to Continue Using Smart Home Devices (t2) | | | |
| H4: Trust Confirmation (t2) | 0.226 | 0.151 | 0.067 |
| **H6: Satisfaction (t2)** | **0.395** | **0.163** | **0.008** |
| DV: Wanting to Block (t2) | | | |
| H3: Trust Confirmation (t2) | -0.024 | 0.275 | 0.466 |
| **H5: Satisfaction (t2)** | **-1.263** | **0.298** | **<0.001** |

*4.2.2 Mediation Analysis.* Figure 4 suggests the possibility of a full mediation, where trust confirmation predicts willingness to block and intention to use variables, but its effect is fully mediated through satisfaction, leading to trust confirmation having non-significant direct effects. Therefore, we conducted a mediation analysis and studied the indirect effects of trust confirmation on these dependent variables. We found that the effect of trust confirmation on wanting to block device communications is fully mediated through satisfaction (indirect effect: -0.521, $p$ = .007), such that if we remove satisfaction from the model, the effect of trust confirmation on wanting to block would be significant ($B$ = –0.387, $p$ < .004). Likewise, the effect of trust confirmation on intention to continue using smart home devices in the future is fully mediated through satisfaction (indirect effect: 0.206, $p$ = .014), such that if we remove satisfaction from the model, trust confirmation significantly predicts intention to continue using smart home devices in the future ($B$ = 0.541, $p$ = .037). Because satisfaction is operationalized here in the context of privacy, the indirect effects operate through participants' evaluation of the benefits and privacy risks of device use. Greater trust confirmation was associated with greater satisfaction, which in turn was associated with stronger continued-use intention and lower willingness to block.

## 4.3 Discussion

In our first study, we investigated the impact of revealing actual data practices to users in their own homes using the IoT Inspector. The report provided information against which participants could evaluate their pre-existing trust expectations. We found that users who had higher trust at the beginning were more likely to have that trust confirmed after viewing the report. This pattern may reflect greater prior familiarity among people who voluntarily signed up to use a monitoring tool, which could have made their trust expectations more likely to align with the practices shown in the report. However,
we did not directly measure participants' prior familiarity or technical knowledge about their devices. This interpretation is therefore tentative and based on the recruitment context rather than directly measured characteristics of the sample.

A key finding specific to this in-situ context is the mediating role of satisfaction. In this sample, the decision to engage in coping behaviors (both passive discontinuance of use and active blocking) was channeled entirely through their satisfaction with data practices. Specifically, when individuals found a violation of their expectations, they felt less satisfied, which in turn lowered their intention to continue using the device. This *satisfaction gatekeeping* likely reflects the real-world stakes of the study: blocking data or discarding a device in a live home environment has tangible functional costs, and users weigh these costs against the benefits before acting. It is also noteworthy that privacy concern was not a significant predictor in this study. One possible explanation is a ceiling effect, given the relatively high mean privacy-concern score of 5.61 on a 7-point scale. However, participants' scores still showed variation, and we did not formally test for a ceiling effect, so this explanation remains tentative. Other possible contributing factors include the small sample size, the use of a single global privacy-concern item, and the characteristics of the self-selected sample.

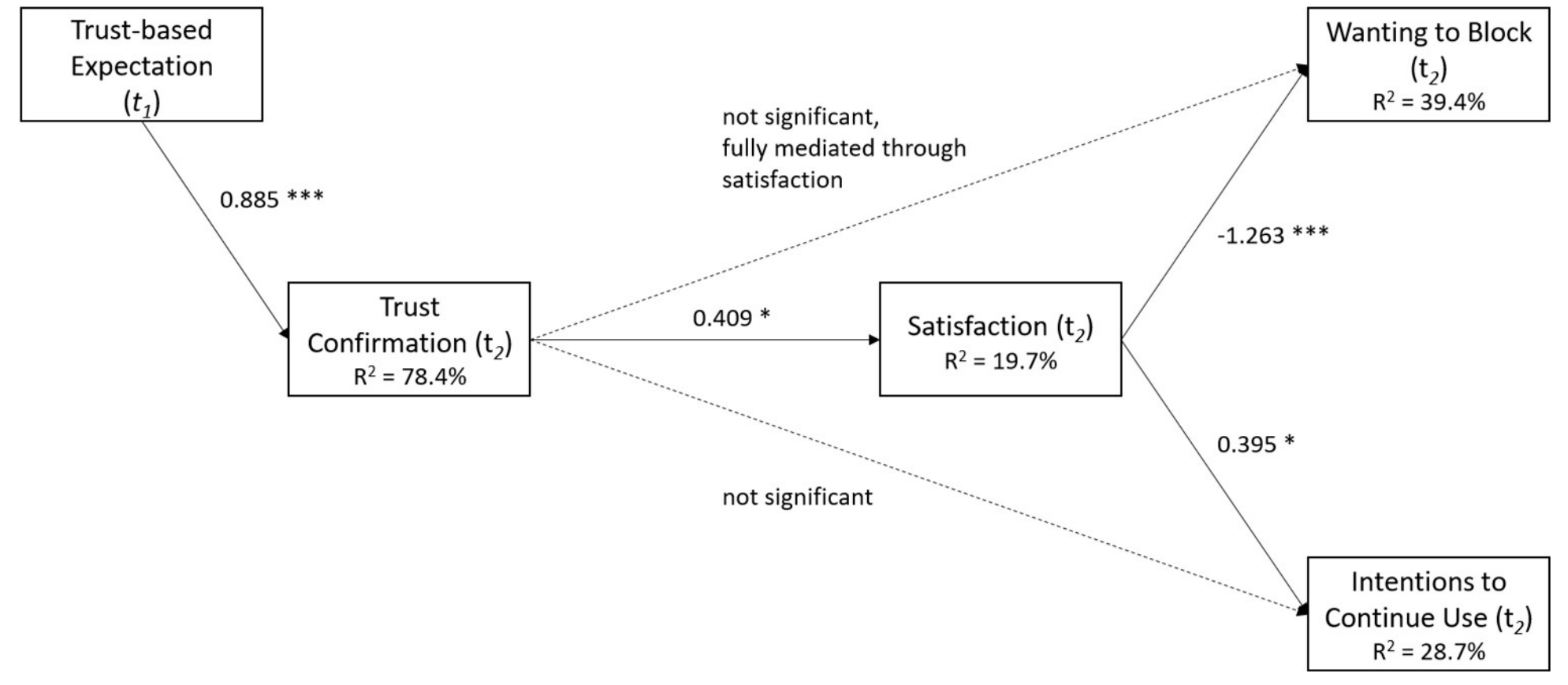

Fig. 4. Satisfaction mediates the effects of trust confirmation on the number of device communications users want to block (indirect effect: -0.521, $p = .007$) and intention to continue using smart home devices (indirect effect: 0.206, $p = .014$).

While providing ecologically grounded evidence, Study 1 faces limitations regarding generalizability. Our participants were self-selected smart-home users who voluntarily installed IoT Inspector and may have been more interested in privacy or more technically motivated than the broader population of smart-home users. Furthermore, the small sample size ($N = 35$) and use of single-item measures limit the strength of the inferences that can be drawn from the path model. In addition, because we did not retain participant-linked report contents, our results should be interpreted as reflecting responses to exposure to a real IoT Inspector report in general, rather than as evidence that any particular report feature (e.g., traffic volume, specific domains) drove participants' post-survey responses. We therefore interpret Study 1 as exploratory and conducted Study 2 with a larger online sample, multi-item measures, and a controlled experimental design to examine whether similar patterns emerged under stronger statistical and measurement conditions.

## 5 Study 2: The Online Experiment

### 5.1 Methods

We conducted a survey-based experiment to examine the privacy expectations and behavioral intentions of smart home owners. The study followed a between-subjects design and was approved by our Institutional Review Board (IRB).

The study consisted of a two-phase Qualtrics survey, as summarized in Figure 5. First, participants identified their most frequently used smart device (smart speaker, smart TV, or smart camera) and specified its brand (e.g., LG, Samsung). They then answered questions regarding their baseline trust, privacy concerns, and usage habits. Subsequently, participants viewed a network activity report that was framed as a real-time analysis of their specific device's traffic. In reality, this was a simulated report dynamically customized to match the user's input from the first survey. To increase the perceived realism and personal relevance of the simulated report, we showed participants an animated progress bar with the message "Analyzing your network data" before showing them this simulated report. Adding this step implied that there is a back-end system processing their device information and that the report is based on real data.

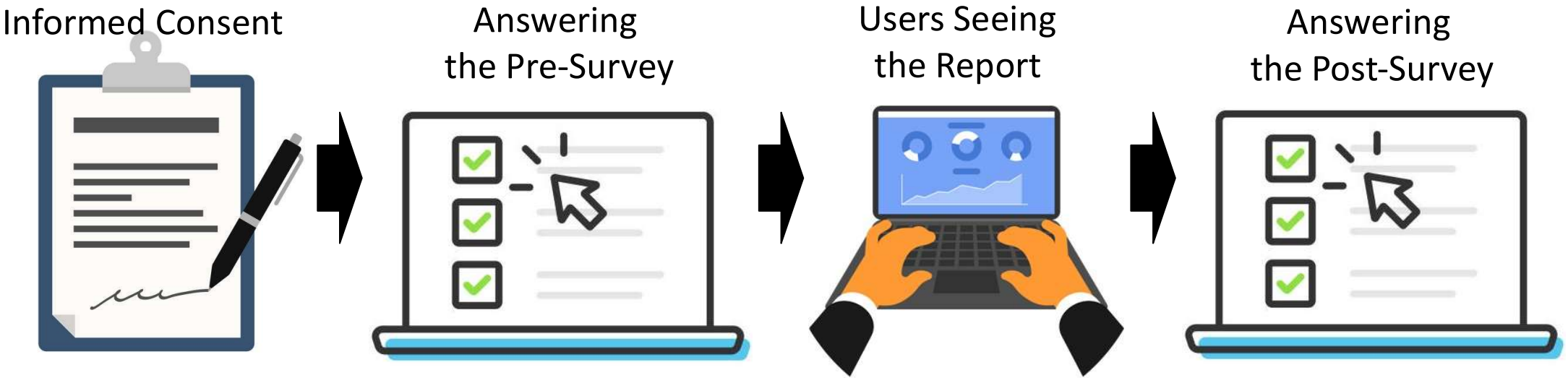

Fig. 5. Participants completed a pre-survey, viewed either a high or low advertising-related network activity report for a device they owned, and then completed a post-survey.

To provide a contextual baseline for comparison, the report displayed traffic volume between the device and its manufacturer (e.g., "LG Server" or "Samsung Server"), representing functionality-related communication, alongside communication with advertising-related domains. The reports were based on a deliberately simplified model in which a smart-home device communicates with manufacturer or functionality-related servers and may also communicate with advertising or tracking domains. This structure is grounded in prior work documenting that some smart devices, particularly smart TVs and smart speakers, communicate with both manufacturer servers and advertising or tracking domains [45, 49]. The manufacturer traffic was held constant across conditions and served as a fixed reference point, while only the volume of advertising-related communication varied. As shown in Figure 6, both reports displayed the same manufacturer traffic (842 KB uploaded and approximately 390 KB downloaded). Advertising-related traffic was 90 KB uploaded and 110 KB downloaded in the *LowAd* condition, compared with 1.2 MB uploaded and 900 KB downloaded in the *HighAd* condition. These volumes were selected to create a controlled experimental contrast and were not calibrated to represent the actual or typical traffic of any particular device or device category.

Upon reviewing the report, participants proceeded to the post-survey to evaluate their experience. First, they assessed whether the reported data practices aligned with their initial expectations. We then measured post-manipulation constructs, including trust in the device, technology self-efficacy, and satisfaction (measured as the perceived trade-off between benefits and privacy risks). The survey also assessed their intention to continue using the device and their general online privacy concerns. Finally, participants responded to open-ended questions regarding their reactions to the report and provided demographic information (e.g., age, education, and income).

*5.1.1 Ethical Considerations and Use of Deception.* Study 2 used temporary deception. Participants were told that the network report reflected a real-time analysis of their own device; in reality, the report was simulated, customized to their stated device, and randomly assigned to the *HighAd* or *LowAd* condition. This design allowed us to control the depicted level of advertising-related communication while increasing the perceived realism and personal relevance of the report by presenting it as reflecting traffic from participants' own devices. The study was reviewed and approved by our Institutional Review Board and was determined to involve minimal risk. Immediately after completing the study, they were informed that the report was simulated and given the option to withdraw their data without affecting their compensation. No participant chose to withdraw.

*5.1.2 Measures.* All multi-item constructs were measured using 7-point Likert scales (1 = Strongly Disagree, 7 = Strongly Agree). With the exception of trust expectation and trust confirmation[2], all constructs were measured both before and after the manipulation to examine possible shifts in participants' attitudes following exposure to the data report.

- Trust-based expectation (pre-survey only): We measured users' expectations of their smart devices prior to the manipulation using the benevolence dimension of trust, which reflects whether a technology acts in the users' best interests [93]. Consistent with Study 1, we adapted this construct to the smart-home privacy context. Participants rated statements such as: "My Smart Device keeps my best interests in mind when handling my personal data."
- Trust Confirmation (post-survey only): After viewing the report, participants rated the extent to which the device's data practices aligned with their expectations. In this case, we followed the literature of measuring confirmation [11]. "The way my smart devices handled my personal data was better than what I expected."
- Satisfaction [95], Intention to Continue Use [11], Technology Self-Efficacy [90], and Collection Concern [70] (Pre and Post): These constructs were measured directly before and after the manipulation to assess whether exposure to the data report shifted participants' evaluations. The extended versions (multi-item) of the same scales used in Study 1 were employed here, with wording adapted to the specific device selected by each participant.

[2] By definition, trust expectation represents beliefs held prior to an event and was therefore measured only in the pre-survey, whereas trust confirmation evaluates the outcome of that event and was uniquely measured in the post-survey

- Awareness (Pre and Post): To assess attentiveness to data handling practices, we included the IUIPC awareness subscale [70] (e.g., the importance of being informed about how personal data is used).
- Unauthorized Use (Pre and Post): We additionally included the IUIPC unauthorized use subscale [70] to assess concerns regarding secondary use of personal data without user consent.

*5.1.3 Qualitative Data:* We collected open-ended responses to better understand participants' expectations and their interpretations of the network-traffic report. Participants were asked whether the data practices of their device aligned with their expectations, and were instructed not to provide any identifying information in this question.

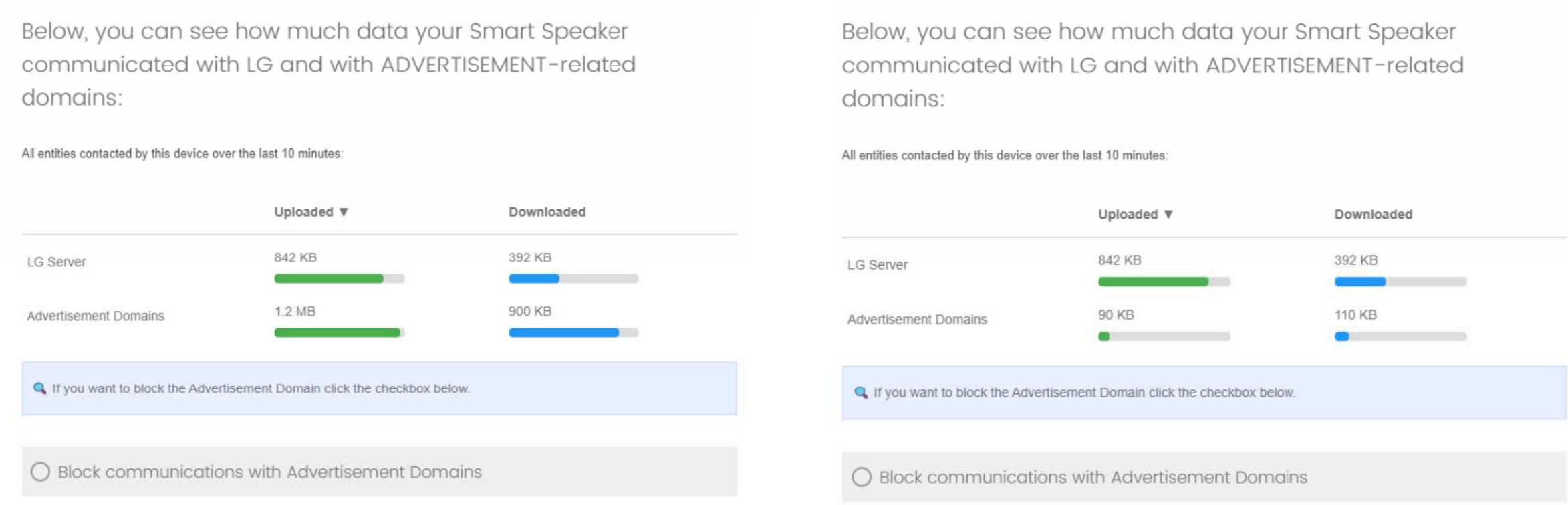

Fig. 6. Simulated network activity reports comparing the HighAd (left) and LowAd (right) conditions. Advertisement traffic volume is displayed relative to the manufacturer baseline (e.g., LG or Samsung servers), with a radio button as a choice option for blocking communications.

*5.1.4 Participant Recruitment.* 132 participants were recruited through Prolific. Prolific is an online platform for recruiting research participants. The website is publicly accessible on the internet. For this experiment, participants had to be 18 years or older, U.S.-based adults, and must use at least one smart home device (smart speaker, smart TV, or smart camera). They received monetary compensation ($4 for the study's duration of approximately 10 minutes) in accordance
with Prolific's guidelines. Only respondents who gave explicit, electronic consent could participate in the study.

*5.1.5 Demographic Information of Participants.* Our participants were, on average, 49 years old. Gender distribution was balanced, with comparable numbers of male (N = 65) and female (N = 63) participants. Two of the participants reported Non-Binary gender, and two preferred not to disclose their gender identity. The majority of participants identified as
White (N = 100), with smaller proportions identifying as Black or African-American (N = 18). Six participants identified as American Indian or Alaska Native, two as Native Hawaiian/Other Pacific Islander, and six selected "Some other race". Among the participants, the education level ranged from high school incomplete or less (N = 1) to a master's/doctorate or other professional degree (medical/law) (N = 31). The most commonly reported level was a four-year college or bachelor's degree (N = 48), followed by some college or an associate's degree (N = 31) and high school graduate/GED (N = 19). Some postgraduate/professional schooling without a postgraduate degree was the least common (N = 2). Participants' annual household income ranged from less than $10,000 (N = 3) to $150,000 or more (N = 15). The most commonly reported income range was $50,000 to less than $75,000 (N = 30), followed by $100,000 to less than $150,000 (N = 25) and $75,000 to less than $100,000 (N = 23). The remaining participants fell into the following income ranges: $10,000 to less than $20,000 (N = 9), $20,000 to less than $30,000 (N = 8), $30,000 to less than $40,000 (N = 11), and $40,000 to less than $50,000 (N = 8). Table 2 reports the full demographic information.

Table 2. Demographic Information of Participants for Study 2 ($N$ = 132)

| Variable | Category | N (%) / M (SD) |
|---|---|---|
| Age (years) | | 49.48 (16.70) |
| Gender | Male | 65 (49.24%) |
| | Female | |
| | | 63 (47.73%) |
| | Non-Binary | 2 (1.52%) |
| | Prefer not to say | 2 (1.52%) |
| | White | 100 |
| | Black or African-American | (75.76%) |
| | American Indian or Alaska Native | 18 (13.64%) |
| Race/ethnicity | | 6 (4.55%) |
| | Native Hawaiian/Other Pacific Islanders | 2 (1.52%) |
| | Some other race | 6 (4.55%) |
| | High school incomplete or less | 1 (0.76%) |
| | High school graduate or GED (includes technical/vocational training) | 19 |
| | Some college or associate's degree | (14.39%) |
| | Four year college degree or bachelor's degree | 31 (23.48%) |
| | | 48 |
| Education | | (36.36%) |
| | Some postgraduate or professional schooling (no postgraduate degree) | 2 (1.52%) |
| | Master's, doctorate, postgraduate or professional degree (medical/law) | 31 (23.48%) |
| | Less than $10,000 | 3 |
| | $10,000 to less than $20,000 | (2.27%) |
| | $20,000 to less than $30,000 | 9 |
| | $30,000 to less than $40,000 | (6.82%) |
| | $40,000 to less than $50,000 | 8 (6.06%) |
| | | 11 (8.33%) |
| | | 8 |
| Income | | (6.06%) |
| | $50,000 to less than $75,000 | 30 (22.73%) |
| | $75,000 to less than $100,000 | 23 (17.42%) |
| | $100,000 to less than $150,000 | 25 (18.94%) |
| | $150,000 or more | 15 (11.36%) |

*5.1.6 Data Analysis:* For the quantitative analysis, we used confirmatory factor analysis (CFA) and structural equation modeling (SEM) to assess pre- and post-survey constructs and to test the hypothesized relationships in our model. All quantitative analyses were conducted in R using packages that are designed for latent variable modeling and regression.

For the qualitative analysis, we employed a thematic analysis approach to examine participants' responses to our open-ended question, in which we asked whether the data practices of their device aligned with their expectations. Following established guidelines for thematic analysis [13], we iteratively coded, compared, and refined themes across the responses to develop a clear understanding of participants' perspectives.

### 5.2 Results

*5.2.1 CFA Report.* A confirmatory factor analysis (CFA) was conducted on both pre- and post-survey data to evaluate the construct validity of the latent variables. The pre-survey model shows an acceptable to good fit across several indices: $\chi^2(209) = 318.9$, $ps < .001$, CFI = 0.956, TLI = 0.947; SRMR = 0.053; and RMSEA = 0.064 [34]. The post-survey measurement model also showed reasonable fit: $\chi^2(231) = 416.7$, $ps < .001$; CFI = 0.937; TLI = 0.925; SRMR = 0.052; and RMSEA = 0.079 [34]. Collectively, these results demonstrate that both CFA models provide support for the measurement structure and justify proceeding to structural modeling analyses. Table 3 represents the survey measures and factor loadings for both surveys. One Trust Confirmation item, highlighted in gray in Table 3, was initially included but had a standardized loading of .543, substantially lower than the loadings of the other Trust Confirmation items. Because its standardized loading was considerably lower than those of the other Trust Confirmation items and its removal improved the measurement model, the item was removed from the final model. The removed item is the same item used as the single-item measure in Study 1. Because it was not retained in the final Study 2 multi-item scale, we acknowledge limited cross-study measurement comparability and avoid treating the two studies as using fully equivalent measures of trust confirmation. Study 1 is therefore interpreted as using a single-item proxy for trust confirmation, whereas Study 2 provides a stronger multi-item measurement test.

Table 4 presents the correlation matrix for the study's principal constructs. The main diagonal values represent the square root of the Average Variance Extracted (AVE) for each construct. In all but one instance, these values exceed the corresponding off-diagonal inter-construct correlations, supporting discriminant validity and indicating that the constructs are distinct. The exception is the strong correlation between Awareness and Unauthorized Use (r=0.844). This high degree of association suggests substantial conceptual redundancy between these two sub-scales. Consequently, we retained Awareness due to its direct relevance to the transparency focus of our study design, while excluding Unauthorized Use from the subsequent analysis.

Table 3. Pre and Post Survey Measurement Items and Factor Loadings

| Subjective Construct | Items | Pre Load | Post Load |
|---|---|---|---|
| Self Efficacy<br>Pre AVE: 0.804<br>Post AVE: 0.793 | I would feel comfortable using my smart home devices on my own. | 0.843 | 0.841 |
| | If I want to, I can use my smart home devices on my own easily. | 0.990 | 0.949 |
| | I would be able to use my smart home devices even if there is no one around to show me how to use it. | 0.848 | 0.877 |
| Collection Concern<br>Pre AVE: 0.761<br>Post AVE: 0.782 | It usually bothers me when online companies ask me personal information. | 0.850 | 0.836 |
| | When online companies ask me for personal information, I sometimes think twice before providing it. | 0.813 | 0.854 |
| | It bothers me to give personal information to so many online companies. | 0.917 | 0.908 |
| | I'm concerned that online companies are collecting too much personal information about me. | 0.905 | 0.936 |
| Satisfaction<br>Pre AVE: 0.754<br>Post AVE: 0.820 | I think my benefits gained from the use of my smart home devices can offset the risks of my information disclosure. | 0.762 | 0.823 |
| | The value I gain from use of my smart home devices is worth the information I give away. | 0.937 | 0.917 |
| | I think the benefits gained from using my smart home devices will be more than the risks of my information disclosures. | 0.897 | 0.970 |
| Intentions to Continue Use<br>Pre AVE: 0.757<br>Post AVE: 0.805 | I intend to continue using my smart home devices rather than discontinue their use. | 0.858 | 0.926 |
| | My intentions are to continue using my smart home devices than use any alternative means. | 0.788 | 0.867 |
| | If I could, I would like to continue my use of smart home devices. | 0.956 | 0.898 |
| Awareness<br>Pre AVE: 0.559<br>Post AVE: 0.583 | Companies seeking information online should disclose the way the data are collected, processed, and used. | 0.743 | 0.866 |
| | A good consumer online privacy policy should have a clear and conspicuous disclosure. | 0.778 | 0.792 |
| | It is very important to me that I am aware and knowledgeable about how my personal information will be used. | 0.720 | 0.610 |
| Unauthorized Use<br>Pre AVE: 0.693<br>Post AVE: 0.721 | Online companies should not use personal information for any purpose unless it has been authorized by the individuals who provided information. | 0.815 | 0.751 |
| | When people give personal information to an online company for some reason, the online company should never use the information for any other reason. | 0.842 | 0.869 |
| | Online companies should never sell the personal information in their computer databases to other companies. | 0.764 | 0.853 |
| | Online companies should never share personal information with other companies unless it has been authorized by the individuals who provided the information. | 0.903 | 0.915 |
| Trust-based Expectation AVE: 0.904 | I expect my smart home devices to put my interest first when handling my personal data. | 0.937 | |
| | I expect my smart home devices to keep my interests in mind when handling my personal data. | 0.954 | |
| | I expect my smart home devices to understand my needs and preferences when handling my personal data. | 0.962 | |
| Trust Confirmation AVE: 0.833 | The way my smart home devices handled my personal data was better than what I expected. | | 0.872 |
| | Overall, most of the expectations regarding how my smart home devices handled my personal data were confirmed. | | 0.543 |
| | My experience confirmed that my smart home devices put my interest first when handling my personal data. | | 0.927 |
| | My experience confirmed that my smart home devices keep my interests in its mind when handling my personal data. | | 0.955 |
| | My experience confirmed that my smart home devices want to understand my needs and preferences when handling my personal data. | | 0.894 |

Table 4. Correlation Matrix. Since Awareness and Unauthorized Use do not meet the discriminant validity criteria, we remove Unauthorized Use from the analyses.

| | Self-Efficacy | Satisfaction | Expectations | Intention | Collection | Awareness | Unauthorized Use |
|---|---|---|---|---|---|---|---|
| Self-Efficacy | 0.896 | | | | | | |
| Satisfaction | 0.256 | 0.868 | | | | | |
| Expectations | 0.139 | 0.268 | 0.951 | | | | |
| Intention | 0.485 | 0.404 | 0.139 | 0.870 | | | |
| Collection | 0.097 | -0.293 | 0.001 | 0.028 | 0.872 | | |
| Awareness | 0.566 | 0.061 | 0.289 | 0.389 | 0.381 | 0.747 | |
| Unauthorized Use | 0.583 | -0.071 | 0.217 | 0.408 | 0.342 | 0.844 | 0.832 |

*5.2.2 Path Model.* The path model (Figure 7) illustrates how trust-based expectations, confirmation, and satisfaction shape users' behavioral intentions. The results of Study 2 are presented in Table 5. This model shows an acceptable fit ($\chi^2$(216) = 269.16, $p$ = 0.008, CFI = 0.981, TLI = 0.977, SRMR = 0.051, RMSEA = 0.044, p = 0.733, CI: [0.023, 0.059]). The results show that pre-survey trust-based expectations significantly predict post-survey trust confirmation ($\beta$ = 0.469, $p < .001$, H1 confirmed), indicating that users with higher trust-based expectations about their smart devices are more likely to report stronger confirmation of those expectations. We also found that stronger trust confirmation is associated with greater satisfaction with the smart device, suggesting that users who feel their expectations are met tend to report higher satisfaction ($\beta = 0.443$, $p < .001$, H2 confirmed). However, trust confirmation does not significantly predict wanting to block ($\beta = -0.226$, $p = 0.059$, H3 rejected) or intentions to continue use ($\beta = -0.037$, $p = 0.619$, H4 rejected). Finally, satisfaction is not a significant predictor of wanting to block ($\beta = 0.140$, $p = 0.319$, H5 rejected). In contrast, satisfaction significantly predicts intentions to continue use ($\beta = 0.546$, $p < .001$, H6 confirmed), such that more satisfied users are more likely to intend to continue using their devices. In addition, we controlled for self-efficacy, awareness, and collection concerns (measured at pre-survey) by regressing them on all of the other variables. We found that awareness negatively predicts trust confirmation ($\beta = -0.218$, $p = .010$). Self-efficacy positively predicts satisfaction ($\beta = 0.236$, $p = .005$) and intentions to continue use ($\beta = 0.312$, $p = 0.032$) while collection concerns negatively predict satisfaction ($\beta = -0.237$, $p = 0.004$). Moreover, collection concerns is the strongest predictor of blocking decisions ($\beta = 0.409$, $p < .001$). This indicates that users with greater collection concerns are substantially more likely to choose to block device communications.

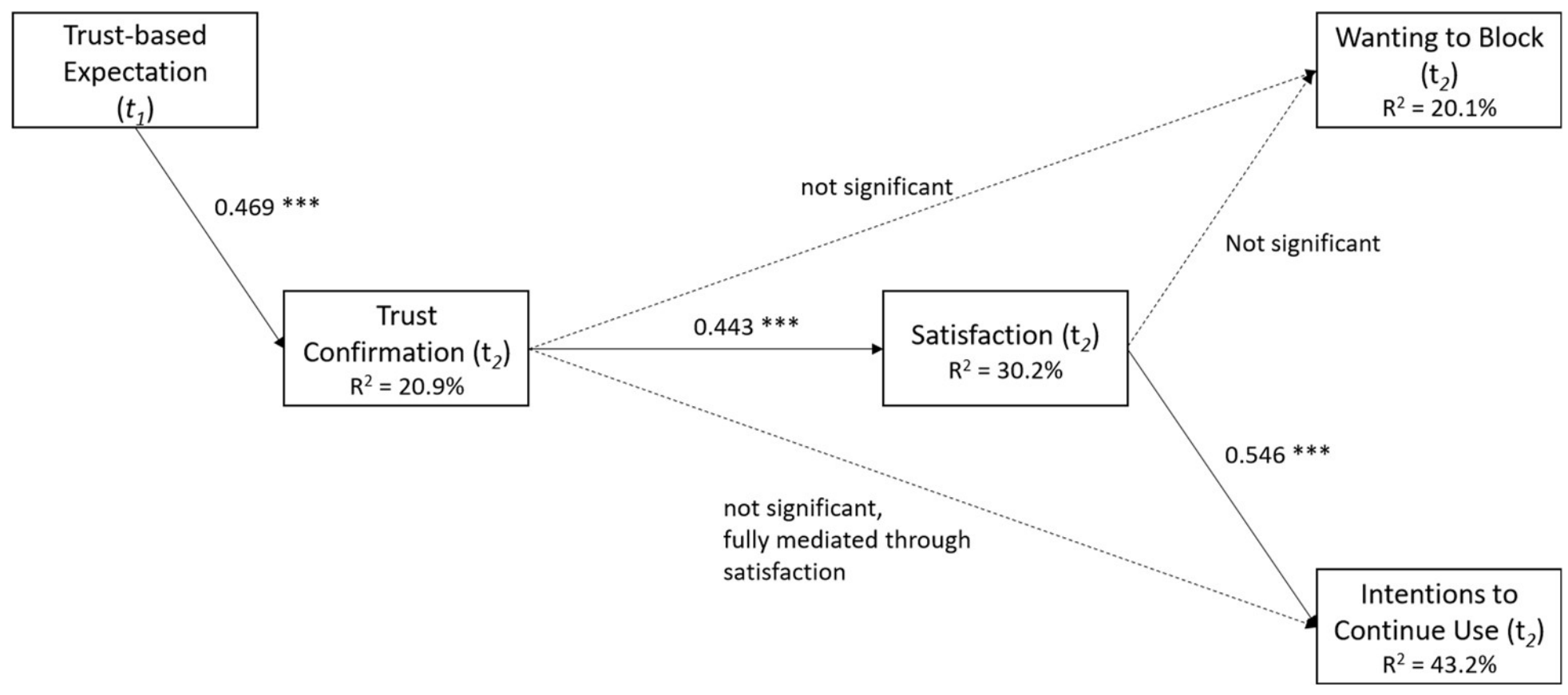

Fig. 7. Study 2 path model illustrating the relationships among key constructs. The model controls for collection and awareness concerns, and privacy self-efficacy by regressing them on all dependent variables; significant effects of these control variables are detailed in Table 5.

Table 5. The results of the path model of study 2. Significant hypotheses are presented in bold format.

| Variable Names | Standardized coefficients | Standard error | p-value |
|---|---|---|---|
| DV: Trust Confirmation (t2) | | | |
| **H1: Trust Expectations (t1)** | **0.469** | **0.116** | **<0.001** |
| Awareness (t1) | -0.218 | 0.095 | 0.010 |
| DV: Satisfaction (t2) | | | |
| **H2: Trust Confirmation (t2)** | **0.443** | **0.103** | **<0.001** |
| **Self-Efficacy (t1)** | **0.236** | **0.100** | **0.005** |
| **Collection Concerns (t1)** | **-0.237** | **0.098** | **0.004** |
| DV: Intentions to Continue Using Smart Home Devices (t2) | | | |
| H4: Trust Confirmation (t2) | -0.037 | 0.088 | 0.619 |
| **H6: Satisfaction (t2)** | **0.546** | **0.118** | **<0.001** |
| **Self-Efficacy (t1)** | **0.312** | **0.188** | **0.032** |
| DV: Wanting to Block (t2) | | | |
| H3: Trust Confirmation (t2) | -0.226 | 0.106 | 0.059 |
| H5: Satisfaction (t2) | 0.140 | 0.109 | 0.319 |
| **Collection Concerns (t1)** | **0.409** | **0.100** | **<0.001** |

Table 6. Regression results predicting post-survey outcomes from pre-survey measures and experimental condition (HighAd vs. LowAd). Each model includes the corresponding pre-measure, condition, and their interaction term. We found significant interaction effects only for Self-Efficacy and Intention to use.

| DV | IVs | Standard coeff | Standard error | $p$-value |
|---|---|---|---|---|
| Post Self-Efficacy | Pre Self-Efficacy | 0.796 | 0.049 | < 0.001 |
| | HighAd (vs. LowAd) | -0.059 | 0.098 | 0.549 |
| | Pre Self-Efficacy × HighAd | 0.291 | 0.099 | 0.003 |
| Post Collection Concerns | Pre Collection Concerns HighAd | 0.894 | 0.039 | < 0.001 |
| | (vs. LowAd) | -0.103 | 0.077 | 0.180 |
| | Pre Collection Concerns × HighAd | 0.027 | 0.079 | 0.736 |
| Post Awareness | Pre Awareness | 0.921 | 0.042 | < 0.001 |
| | HighAd (vs. LowAd) | 0.023 | 0.079 | 0.762 |
| | Pre Awareness × HighAd | -0.077 | 0.086 | 0.373 |
| Post Unauthorized Use | Pre Unauthorized Use HighAd | 0.916 | 0.038 | < 0.001 |
| | (vs. LowAd) | -0.032 | 0.073 | 0.655 |
| | Pre Unauthorized Use × HighAd | -0.082 | 0.076 | 0.281 |
| Post Satisfaction | Pre Satisfaction | 0.850 | 0.051 | < 0.001 |
| | HighAd (vs. LowAd) | 0.147 | 0.098 | 0.137 |
| | Pre Satisfaction × HighAd | -0.161 | 0.103 | 0.119 |
| Post Intention | Pre Intention | 0.801 | 0.051 | < 0.001 |
| | HighAd (vs. LowAd) | 0.147 | 0.100 | 0.145 |
| | Pre Intention × HighAd | -0.212 | 0.103 | 0.042 |

*5.2.3 Post-hoc Analyses with Experimental Manipulations.* To investigate whether the experimental manipulation (HighAd vs. LowAd) changed user attitudes, we conducted post-hoc linear regressions predicting each post-survey construct using its corresponding pre-survey baseline, the experimental condition, and their interaction term (Table 6). Across all constructs, pre-survey measures strongly predicted their respective post-survey outcomes. Significant interactions between pre-survey scores and the experimental condition were observed only for Self-Efficacy ($p = .003$) and Intention ($p = .042$), indicating that the experimental manipulation specifically moderated changes in these variables (visualized in Figure 8). In contrast, no significant interactions were found for Collection Concerns, Awareness, or Satisfaction, suggesting the experimental manipulation did not change these attitudes beyond baseline levels. [3]

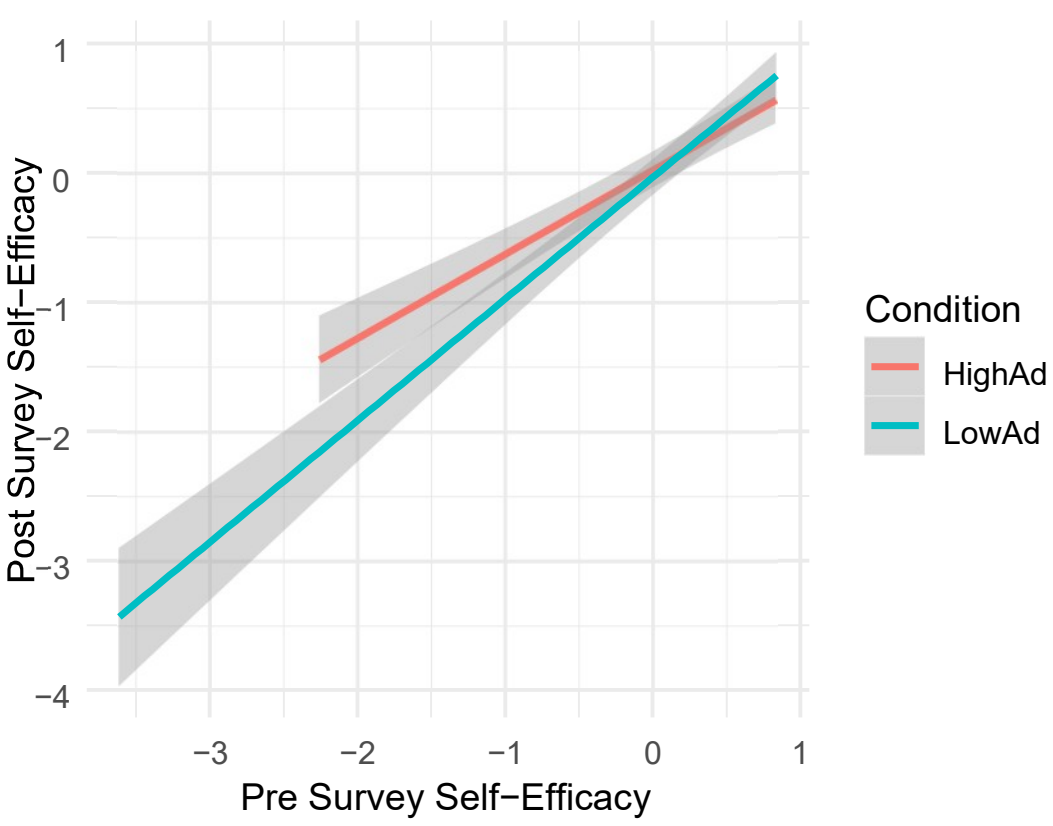

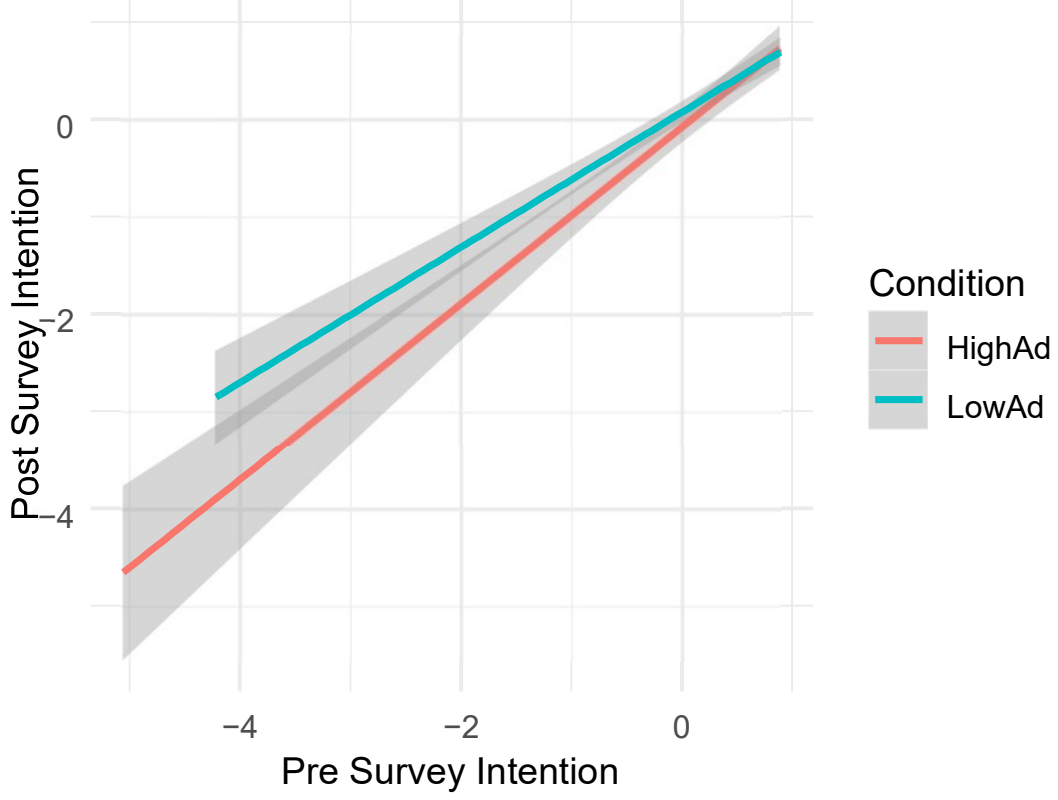

(a) Pre–Post Self-Efficacy by condition

(b) Pre–Post Intention by condition

Fig. 8. Pre–post survey relationships between self-efficacy and intention to continue using across HighAd and LowAd conditions.

*5.2.4 Qualitative Insights.* To complement the quantitative results of the path model, we conducted a qualitative thematic analysis on participants' open-ended responses to the question: "Were the data practices of your smart home devices (Smart TV, Smart Speaker, and Smart Camera) consistent with your expectations?" Of the 132 participants, 130 responded to this open-ended question; two participants left the question blank. This resulted in 65 responses from each experimental condition. Participants answered this question after viewing the simulated network-activity report. Their responses should therefore be interpreted as reactions to the data practices depicted in the report, rather than as assessments of their devices' actual data practices. Our analysis revealed three primary themes: *Data Practices Consistent with Expectation*, *Distrust Toward Data Practices*, and *Others* (Table 7). The *Others* category encompasses responses that were neutral, unrelated to specific data practices, or no-expectation responses. Collectively, these themes capture participants' interpretations and reactions to the data practices depicted in the simulated report. Although the same themes appeared in both conditions, their descriptive distributions differed, particularly for expressions of inconsistency or distrust.

The proportion of participants who described the report as consistent with their expectations was similar across conditions (HighAd: 39 of 65, 60.0%; LowAd: 38 of 65, 58.5%). However, expressions of inconsistency or distrust were descriptively more common in the HighAd condition (12 of 65, 18.5%) than in the LowAd condition (7 of 65, 10.8%). For example, one participant in the HighAd condition stated, "The data practices were not consistent with my expectations. I was surprised by how much data was sent to advertisement-related domains compared to the actual TV manufacturer." In contrast, some participants in the LowAd condition described the depicted volume of advertising-related communication as lower than they had expected and expressed reassurance. These findings indicate that an overall judgment of consistency did not necessarily capture whether participants approved of, trusted, or felt comfortable with the depicted practices.

*5.2.5 How Participants Make Judgments of Consistency.* A majority of participants in both conditions described the depicted data practices as consistent with their expectations. Within these responses, we also identified four interpretive sub-themes that reflected how participants arrived at this judgment: Normalization or pragmatic acceptance, priorknowledge confirmation, favorable deviation from expectations, and partial surprise within an overall judgment of consistency. Responses could reflect more than one sub-theme. Responses that were too brief to determine how participants arrived at their judgment were not assigned an interpretive sub-theme.

Table 7. Frequencies of qualitative themes by experimental condition, with representative examples. Responses could reflect more than one subtheme; therefore, subtheme frequencies may exceed the corresponding parent-theme totals.

| | Frequency ([illegible]) | | |
|---|---|---|---|
| Theme | HighAd | LowAd | Representative Example |
| Data Practices Consistent with Expectation | 39 | 38 | |

[3] The main study did not include an internal manipulation-check item assessing participants' perceived level of advertising-related activity. We therefore conducted a supplementary manipulation-check study showing participants the HighAd/LowAd reports identical to the main study. After browsing the report, participants answered the question "Based on the report you just reviewed, how much data do you feel your Smart TV shared with advertisingrelated companies?" (rated from "None at all" to "A very large amount"). Using a paired t-test, participants perceived significantly more advertising-related activity in the HighAd report than in the LowAd report ($M = 3.97$ vs. 2.67, $t(29) = 7.48$, $p < .0001$), suggesting that the stimuli differed perceptibly in the intended direction. Because this check was conducted in a separate sample, it does not establish how participants in the main study perceived the manipulation.

| | | | |
|---|---|---|---|
| Normalization or pragmatic acceptance | 11 | 4 | “yes, I think that all smart devices are using and abusing our data. We are almost never given a choice but to accept their terms of service. I also believe that my smart phone is listening to my conversations.” |
| Prior-knowledge confirmation | 23 | 18 | “yes, The data practices of my Smart TV matched my expectations, as I was aware it would collect viewing history, app usage, and voice commands for recommendations and ads, and I found the expected privacy settings during setup.” |
| Favorable deviation from expectations | 4 | 16 | “Yes, the levels of information/advertisement sharing were actually a bit lower than my expectations.” |
| Overall consistency with partial surprise | 5 | 1 | “The data practices of my Smart Speaker were somewhat consistent with my expectations. I anticipated that voice recordings might be stored and analyzed to improve performance, but I was surprised by how much data could be collected passively, such as background noise or frequent usage patterns. I would prefer more transparency and easier control over privacy settings.” |
| Distrust Toward Data Practices | 12 | 7 | |
| Volume exceeded expectations | 9 | 5 | “No, because I have no idea why my Echo would be transferring so much (or really any) data just in the last 10 minutes.” |
| Protection or consent violated | 4 | 2 | “No. I expected my smart TV to protect my data, if they were to use it in other ways I expect to have been asked for permission.” |
| Others | 14 | 20 | “I didn’t really have any expectations.” |

Normalization or pragmatic acceptance. Some participants described the depicted practices as expected because they viewed data sharing as routine or unavoidable in smart-device ecosystems ($n$ = 15; HighAd: $n$ = 11, LowAd: $n$ = 4). For these participants, consistency reflected resignation to an expected practice rather than approval or satisfaction. "yes, I think that all smart devices are using and abusing our data. We are almost never given a choice but to accept their terms of service. I also believe that my smart phone is listening to my conversations."

Prior-knowledge confirmation. Other participants referred to specific practices they already expected, such as the collection of viewing history, app usage, or voice-command data ($n$ = 41; HighAd: $n$ = 23, LowAd: $n$ = 18). In these responses, the report appeared to confirm participants’ prior understanding of smart-device data practices. "yes, The data practices of my Smart TV matched my expectations, as I was aware it would collect viewing history, app usage, and voice commands for recommendations and ads, and I found the expected privacy settings during setup."

Favorable deviation from expectations. Some participants, particularly in the LowAd condition, stated that the report showed less advertising-related communication than they had anticipated ($n$ = 20; HighAd: $n$ = 4, LowAd: $n$ = 16). Although these responses had initially been categorized as broadly consistent with expectations, their explanations reflected relief that the depicted practices were less extensive than expected. "Yes, the levels of information/advertisement sharing were actually a bit lower than my expectations."

Overall consistency with partial surprise. Some participants, more frequently in the HighAd condition, characterized the report as generally consistent with their expectations while also identifying particular practices that surprised or concerned them, such as communication with third-party advertising domains ($n$ = 6; HighAd: $n$ = 5, LowAd: $n$ = 1). These responses show that an overall judgment of consistency could coexist with concern about specific communications. "The data practices of my Smart Speaker were somewhat consistent with my expectations. I anticipated that voice recordings might be stored and analyzed to improve performance, but I was surprised by how much data could be collected passively, such as background noise or frequent usage patterns. I would prefer more transparency and easier control over privacy settings."

A smaller set of participants reported that the depicted practices did not match their expectations, expressing discomfort, suspicion, or a sense that the device was doing more than they had assumed (HighAd: n = 12, 18.5%; LowAd: n = 7, 10.8%). These expressions were descriptively more common in the HighAd condition, consistent with the manipulation. Participants developed distrust for two main reasons: concern about the amount of data leaving the device, and concern about a perceived breach of protection or consent norms.

Volume exceeded expectations. Some participants located their distrust in the sheer amount of communication depicted, rather than any specific recipient or purpose (n = 14; HighAd: n = 9, LowAd: n = 5), often expressing surprise that the device was communicating at all in a short window. "No, because I have no idea why my Echo would be transferring so much (or really any) data just in the last 10 minutes." This mirrors the quantitative finding that collection concerns, rather than satisfaction, predicted willingness to block in Study 2.

Protection or consent violated. Some participants distrusted the device because they believed it should protect their data or ask permission before using it in other ways (n = 6; HighAd: n = 4, LowAd: n = 2). Their concern was about whether the device followed appropriate privacy norms, not simply how much data it shared. This concern appeared even when the amount of reported traffic was relatively low. As one participant explained, “No. I expected my smart TV to protect my data. If they were to use it in other ways, I expected to be asked for permission.”

Although the same themes appeared in both conditions, participants reached their judgments of consistency for different reasons. HighAd participants more often described the practices as normal or unavoidable and were more likely to express surprise or concern about specific practices, whereas LowAd participants more often expressed reassurance that the report showed less data sharing than they had expected. These findings show that consistency with expectations did not uniformly indicate trust, approval, or satisfaction. The same overall judgment could reflect informed prior knowledge, resignation to practices perceived as unavoidable, relief that the depicted communication was less extensive than anticipated, or continued concern about particular data flows. The qualitative findings, therefore, complement the path model by showing that quantitatively similar levels of confirmation may arise from different forms of reasoning and evaluation.

## 5.3 Discussion

Study 2 extends our investigation of smart-home privacy expectations to a broader sample by experimentally manipulating the level of advertising-related communication. Consistent with Study 1, we find that trust-based expectations strongly predict trust confirmation, and trust confirmation positively predicts satisfaction. Across both studies, satisfaction remains the strongest predictor of intention to continue using the device, reinforcing the core ECT pathway that users’ willingness to stay with a smart-home device depends largely on whether the benefits they perceive from the overall experience outweigh the risks involved. We also found that awareness, operationalized as endorsement of transparency principles, was associated with lower trust confirmation. The items in this construct assessed whether participants believed companies should clearly disclose how personal information is collected and used, and whether being informed about such practices was important. One possible explanation is that participants with stronger expectations for transparency evaluated the advertising-related communications depicted in the report more critically, resulting in lower trust confirmation. This interpretation remains tentative because we did not directly measure how participants applied these transparency expectations when evaluating the report.

However, Study 2 revealed a critical nuance regarding active coping behaviors. Unlike the first study, where satisfaction fully mediated willingness to block, Study 2 participants’ willingness to block device communications was driven primarily by collection concerns rather than satisfaction. This pattern is consistent with a more immediate concern-based response in Study 2, whereas the Study 1 findings are consistent with a broader evaluation of the privacy risks and benefits of device use after participants viewed their actual network traffic. One possible explanation for this divergence is the composition of the two samples. Study 1 participants were self-selected users who had joined the IoT Inspector waiting list and installed the tool, whereas Study 2 drew from a broader online sample. However, because we did not measure participants’ motivations, technical expertise, or privacy motivation, we cannot determine whether the samples differed on these characteristics or whether sample composition explains the distinct patterns. Other differences between the studies may also have contributed. Study 2’s simulated report, despite being framed as reflecting the participant’s own device, may have provided less real-device salience than Study 1, where participants viewed reports generated from their own devices’ actual, ongoing traffic. Relatedly, the simulated, one-time nature of Study 2’s between-subjects design may have made blocking feel less consequential than in Study 1’s field deployment, leading participants to rely more on general collection concerns than on a satisfaction judgment tied to the specific report. We therefore present sample composition, real-device salience, and the perceived consequences of blocking as plausible but untested explanations for the divergence.

Furthermore, our experimental manipulation provided insight into how the depicted volume of advertising-related communication influenced user outcomes. The condition moderated the relationships between participants’ baseline and post-survey self-efficacy and continued-use intentions, indicating that the HighAd and LowAd reports affected participants differently depending on their baseline levels. This underscores that transparency alone is not neutral; the content of the report, specifically the depicted volume of advertising-related communication, can shape users’ self-efficacy and continued-use intentions differently depending on their prior states. We emphasize, however, that the LowAd and HighAd reports represented experimentally manipulated levels of advertising-related communication, not objectively “good” or “bad” privacy practices, and our design does not establish which level, if either, manufacturers should regard as acceptable.

# 6 General Discussion

Our research integrates findings from an in-situ field study ($N$ = 35) and a controlled experiment ($N$ = 132) to demonstrate how inducing transparency influences user attitudes and behaviors. By applying Expectation-Confirmation Theory (ECT) to the domain of smart home privacy, we show that confirming trust expectations is pivotal for maintaining user satisfaction. When transparency tools reveal unexpected data practices, satisfaction erodes, jeopardizing the user’s intent to keep the device. Below, we discuss the theoretical and practical implications of these findings.

## 6.1 Theoretical Implications

The Expectation-Confirmation Theoretical framework examines users’ intentions to continue using a system as a function of the extent to which their expectations of the system are confirmed. We conceptualize users’ attempts to remedy unconfirmed expectations as corrective coping behaviors and note that lowering intentions to continue using the system is only one such response. As an initial examination of an active, control-oriented response, we studied whether users were willing to use an in-tool control to mark unfavorable device communications they wished to block in the future. We therefore frame the extension of ECT to active coping as promising, though not yet established. Establishing when and how trust confirmation predicts active coping will require examining a broader range of behaviors, such as actually blocking traffic, changing device settings, disabling features, deleting collected data, or discontinuing specific data flows, across more diverse populations. Users may also employ other coping behaviors, such as speaking quietly around smart speakers or providing them with misleading information [68].

Furthermore, we specifically studied privacy-related expectations. Previous studies in the privacy literature that adopted the ECT framework measured confirmation in broad terms, such as whether the functionality is, overall, better than users’ expectations [40, 92]. Although there is a trade-off between privacy and functionality in many scenarios [91, 95], privacy and functionality are two distinct concepts. Therefore, by studying the confirmation of functionality expectations, we will not be able to examine users’ privacy expectations and the consequences of not confirming them. We specifically studied users’ trust expectations of how they expect their devices to handle their data. This enabled us to draw the consequences of confirming or rejecting users’ trust expectations. While ECT only considers one type of expectation, future studies can extend this framework by considering several parallel expectations and understanding the role of each expectation in user satisfaction and device usage.

We conducted a mediation analysis that showed satisfaction can fully mediate the effects of trust confirmation on coping behaviors (i.e., intention to use) in both studies. This is an important finding for the ECT literature, as it suggests that confirmation can lead to behavioral effects, sometimes solely through satisfaction, improving our understanding of the ECT’s applicability to real-world scenarios. It is important to note that the satisfaction measured in our studies was not general satisfaction with the device, but an appraisal of whether the benefits of device use outweighed the risks of information disclosure. This construct overlaps conceptually with perceived disclosure value and privacy calculus. Accordingly, alternative theoretical interpretations are plausible, and ECT should be understood as one theoretically grounded explanation of the observed pattern rather than the only possible explanation.

## 6.2 Practical Implications

Our research has important implications for smart home manufacturers and service providers. We studied transparency after adoption, as participants were already using smart-home devices, and found that when the data practices made visible did not align with users' prior trust-based expectations, users reported weaker intentions to continue using the device. This does not imply that data sharing itself necessarily reduces continued-use intentions; rather, the mismatch between users' expectations and the practices made visible appears to matter. Manufacturers should therefore not assume that purchasing and using a product means that users understand, accept, and are satisfied with its data practices.

To improve user satisfaction, companies should explore means to promote transparency post-purchase when a device is already in use. We navigated one of these ways by providing users with a report on the data traffic of their network. While this approach to inducing transparency damaged users' satisfaction with the devices, we should consider that the tools we used for promoting transparency, IoT Inspector and the report on Qualtrics, were generated by third-party. The situation will arguably be different if the devices themselves promote this transparency and bring any potentially unfavorable data practices to the user's attention. Research shows that pointing out a system's flaws by the provider can even improve users' trust in that provider [57]. Therefore, devices can check in with users by providing them with a report of their data communications without compromising their trust.

Additionally, when a device's privacy practices do not align with the user's personal preferences, the device should provide means to bridge this gap (e.g., through privacy settings). Providing a control mechanism is especially important, as some users may otherwise reduce their intentions to use the devices. Our two studies also suggest that the drivers of active control are not uniform across populations and settings, which bears on how such controls should be designed and surfaced. In the in-situ field study, willingness to block was associated with satisfaction, whereas in the controlled experiment, it was driven primarily by collection concerns. This divergence implies that a blocking control may be reached for as part of an overall cost-benefit evaluation of the device by some users, but as a more direct response to data-collection concern by others, and manufacturers should not assume a single motivational pathway when deciding how to present privacy settings. Our work examined two coping behaviors that users may employ when encountering undesirable data practices: reducing intentions to continue using the device and seeking to limit undesirable device communications. The former is a passive coping behavior, preventing users from benefiting from their smart home devices altogether, and should arguably be the last resort. The latter is an active coping behavior, allowing users to continue benefiting from the devices, at least to some degree. We can implement procedures that allow active coping strategies in various ways. For example, in the case of an unfavorable communication, the user and device can engage in a dialogue where the device informs the user about the implications of a change in the communication settings for offered services (e.g., whether a service will be compromised or interrupted by withholding data, and if there are alternatives). This procedure could give users greater control over their data and support trust, as perceived control is associated with trust [42]. This dialogue-based approach also illustrates what granular controls could mean in practice. Granular controls would allow users to act on specific communications rather than choose between accepting all data sharing and discontinuing use of the device. Our study materials illustrated this approach by separating manufacturer-related from advertising-related communications and allowing participants to mark particular communications for future blocking. For such controls to be accessible, the dialogue should describe the available choices and their likely consequences in plain language, including whether blocking a communication may affect device functions or services.

While we highlighted the benefits of implementing control mechanisms, we would like to emphasize a major setback: exercising control is mentally demanding [14, 37]. The situation is arguably worse for smart home environments with various smart devices, as controlling the data practices of each device is challenging, time-consuming, and not feasible [77]. Our studies examined only a simple control that allowed participants to mark particular communications for future blocking. More complex controls across multiple devices may impose additional interpretation and management demands, reinforcing the importance of designing simplified and, where appropriate, partially automated privacy controls. Scholars explored various ways to simplify exercising control, such as designing dialogue-based systems that allow users to create security rules using natural language [37]. To further simplify user interaction with privacy controls and ease the management burden on smart home device users, we should explore ways to complement simplified control mechanisms with some level of automation. For example, the system can learn users' privacy preferences over time and apply these preferences to new devices.

The qualitative findings further suggest that users who report that the depicted practices were consistent with their prior expectations may nevertheless benefit from different forms of support. For example, participants who reported overall consistency but expressed surprise about particular communications may benefit from controls that allow them to act on those specific data flows. Those whose responses reflected normalization or pragmatic acceptance may instead benefit from prompts that encourage them to consider whether the data practices they expect are also acceptable to them and what they could do about them. Transparency and control mechanisms should therefore account for the different ways users interpret the same information rather than providing the same notice to everyone.

### 6.3 Methodological Implications: The Cost of Ecological Validity

Studying real-world scenarios and moving beyond convenience sampling methodologies is crucial in understanding users' attitudes and behaviors towards smart home devices. Emami-Naeini et al. [25] interviewed individuals who had purchased IoT devices and found that privacy and security were not a major consideration for most participants at the time of purchase. However, in another study, they surveyed individuals about smart devices from MTurk, a crowdsourcing platform, survey respondents showed higher interest in their willingness to purchase IoT devices if they knew their data would not be retained or shared with others [24]. While these two studies have valuable implications for marketing, they also highlight the importance of in-situ studies and the fact that real-life scenarios with real users may have different findings and implications than those with individuals from crowdsourcing platforms or non-users [7, 78]. Therefore, we call for more in-situ studies on privacy decision-making, especially in the context of smart homes. However, a major challenge in such studies is recruiting participants since researchers may not have access to a large sample as they do in convenience sampling methodologies. In Study 1, participants were self-selected smart-home users who had voluntarily signed up to use IoT Inspector and may therefore have differed from the broader population of smart-home users. Recruiting more people from this demographic was not possible, as we had already contacted everyone who had signed up to use the IoT Inspector, highlighting one of the recruitment difficulties of such in-situ studies.

Taken together, the two studies offer complementary methodological strengths. In Study 1, participants viewed reports generated from their own smart-home devices' network traffic in their homes, providing ecologically grounded evidence of responses to real device communications, but the design did not allow us to isolate the effects of particular report features. Study 2 instead used simulated reports in a controlled online experiment, allowing us to manipulate the depicted level of advertising-related communication, but not to make claims about participants' actual device behavior. Future work should combine these strengths by connecting users' prior expectations, their devices' measured communications, and their subsequent perceptions and actions within a single longitudinal design.

## 7 Limitations and Future Work

While our in-situ methodology has several unique advantages, such as studying real smart home device users in their own environment and not having incentive-induced bias [43], it has important limitations. Our results should not be generalized to the broader population of smart home device users. Our participants voluntarily installed and used the IoT Inspector. This self-selection may mean that participants were particularly interested in learning about their devices' data practices and differed from the broader population of smart-home users. However, because we did not measure their motivations, technical literacy, or prior familiarity with smart-device data practices, we cannot determine the nature or extent of these differences. These unmeasured characteristics may also have shaped participants' initial expectations and interpretations of the reports. Additionally, we do not have demographic data such as age, gender, and education status for our participants. In line with our IRB guidelines, we opted not to collect demographic data to minimize the risk of de-anonymization. Furthermore, asking for demographic data could demotivate IoT Inspector users from participating in our study, lowering our response rates. Moreover, as participants joined the study on a voluntary basis and without monetary compensation, we minimized the survey questions. Consequently, Study 1 relied on single-item measures for each construct. This prevented us from assessing internal reliability or discriminant validity, and the path model should therefore be interpreted as exploratory rather than as a confirmatory test of the underlying constructs. Study 2 partially addresses this limitation through its larger sample and multi-item measures validated using confirmatory factor analysis.

A further limitation of Study 1 is that we did not retain report features in a participant-linked form. Although participants viewed reports generated from their own device traffic, we could not examine how specific features, such as traffic volume, destination domains, or third-party recipients, related to their expectations and post-survey responses. Study 1 should consequently be interpreted as examining responses after exposure to a real network-activity report, rather than identifying which report contents drove those responses. Future work should link participants' report features with their pre- and post-survey measures.

Our second study, the online experiment, faced a limitation regarding the scope of its manipulation. We focused specifically on the volume of advertising-related data traffic to test user reactions. However, privacy decision-making is complex and multifaceted; users likely consider not just the amount of data being shared, but the specific nature of that data (e.g., sensitive audio recordings versus routine usage logs) and the reputation of the specific third-party entities involved. By isolating traffic volume as the primary independent variable, our study may overlook how different types of data sensitivity interact with trust to influence blocking decisions. Future research should explore these nuances by manipulating data types and sensitivity levels alongside traffic volume. Study 2 also did not embed an internal manipulation-check item; we addressed this with a separate supplementary check confirming that the conditions differed in the intended direction, though embedding such a check within the main protocol would be preferable in future work. Relatedly, our survey items referred to "information disclosure" generally rather than identifying specific recipients. Future work could study data types and recipients, such as manufacturers, advertisers, or other third parties, to examine how these factors shape privacy judgments and intended responses. More broadly, because the two studies differed in sample composition, study setting, and whether the reports reflected real or simulated device traffic, we cannot determine which of these differences accounts for the association of willingness to block with satisfaction (in Study 1) and with collection concern (in Study 2). Future work should vary these factors more systematically to examine when each pathway emerges.

Another limitation of our study is that we cannot draw conclusions about specific devices individually (e.g., smart speakers, smart TVs). For example, we are not able to pinpoint whether violations of users' expectations by a specific smart home device, through poor communication or concealing data practices, can influence users' perceptions of other devices or if violations by some types of devices have more detrimental effects on users' trust and usage intentions. These are important areas for future research as they have policy implications. Suppose bad actors can easily compromise users' trust in other smart home devices. In this case, it would be even more critical to establish industry-wide standards on transparency and stricter control mechanisms to monitor alignment between devices' data practices and their privacy policies. Additionally, policymakers can propose legislation imposing penalties on manufacturers and service providers who broadly violate users' expectations. Furthermore, since users' security and privacy attitudes towards smart devices are context-dependent and evolve as they acquire more information [64], future research should examine the implications of adaptive legislation that evolves with technological advancements and user feedback.

Moreover, future research can explore whether the changes we observed in this study are long-term or short-lived. In case of a long-term change, smart home devices may need to regain their users' trust prior to their satisfaction and usage, while a short-term change may revert when users resume using their devices and get immersed in their experiences. A long-term study is more important when we consider disclosure decisions not only as a function of privacy preferences but also as a function of disclosure benefits. We studied expectation confirmation and satisfaction in the context of devices' data practices. Device performance is arguably a different domain. If users are so reliant on a technology to which they need to disclose data to make it functional, then being aware of the privacy practices of that device may not lead to coping behaviors. For example, individuals who want to navigate using the GPS will have to let go of their physical location. Future studies should explore the interplay between performance and functionality expectations and trust and privacy expectations in the context of smart home devices.

Finally, our studies focused on primary device users who owned, configured, or administered their smart-home devices. However, smart-home devices often operate in shared spaces where household members, guests, and other bystanders may also be subject to data collection without comparable awareness of, or access to, device controls [33, 98]. Transparency and control mechanisms designed for primary users may therefore not adequately protect individuals who lack the access or authority to view device communications or modify privacy settings. Future work should examine how transparency mechanisms can inform all affected individuals and how privacy controls can accommodate differing preferences and power relationships among device owners, household members, guests, and other bystanders.

## 8 Conclusion

This research bridges a critical gap in smart home privacy by integrating findings from an in-situ field study and a controlled experiment to demonstrate that transparency is a double-edged sword: confirming trust reinforces continued use, while unexpected data practices erode the user-device relationship. Our application of Expectation-Confirmation Theory reveals that satisfaction predicts retention, whereas specific privacy risks trigger active defensive measures, suggesting that the current *notice and consent* model is insufficient. Instead, manufacturers must move beyond simple transparency to provide granular, accessible controls that empower users to align device behavior with their privacy norms, ensuring a smart home ecosystem built on verified trust rather than blind acceptance.

## Acknowledgments

This research was supported in part by funding from Google through the CyberNYC initiative. The sponsor had no influence on the study design, data analysis, interpretation, or reporting.

# A Appendices

## A.1 Full Path Model For Study 2

| Variable Names | Standardized coefficients | Standard error | p-value |
|---|---|---|---|
| DV: Confirmation (t2) | | | |
| Privacy Expectations (t1) | 0.541 | 0.375 | 0.149 |
| Privacy Concerns (t1) | 0.096 | 0.186 | 0.605 |
| DV: Satisfaction (t2) | | | |
| Perceived Benefits (t1) | 0.098 | 0.184 | 0.596 |
| Perceived Risks (t1) | -0.236 | 0.210 | 0.260 |
| **Confirmation of Privacy Beliefs (t2)** | **0.518** | **0.213** | **0.015** |
| Privacy Concerns (t1) | 0.172 | 0.177 | 0.329 |
| DV: Intentions to Use IoT (t2) | | | |
| Intentions to Use IoT (t1) | 0.287 | 0.176 | 0.104 |
| **Satisfaction (t2)** | **0.469** | **0.111** | **0.001** |
| Confirmation of Privacy Beliefs (t2) | 0.221 | 0.155 | 0.153 |
| Privacy Concerns (t1) | -0.019 | 0.144 | 0.894 |
| DV: Privacy Feature Use (t2) | | | |
| **Satisfaction (t2)** | **-0.826** | **0.225** | **<0.001** |
| Confirmation of Privacy Beliefs (t2) | 0.003 | 0.054 | 0.951 |
| Privacy Concerns (t1) | 0.058 | 0.050 | 0.252 |

Table 8. The non-significant effects were trimmed out and the final model only constitutes the significant effects, presented in bold style. However, this table also shows the coefficient and p-values for the trimmed effects before they were trimmed out.

## A.2 Study 1 Instrument

Study 1 used single-item measures for each construct in order to minimize participant burden in an uncompensated field deployment (see Section 4.1.2). For each construct, we selected either the highest-loading item from the original validated instrument or, based on discussion among the authors, the item that most directly captured the construct's conceptual definition. All items were rated on a 7-point Likert scale ranging from “Strongly Disagree” (1) to “Strongly Agree” (7).

Participants completed the pre-survey immediately after providing informed consent and before viewing the IoT Inspector report. After interacting with the report for at least five minutes, they were prompted to complete the post-survey. Figure 9–14 show the IoT Inspector interface that Study 1 participants interacted with, in the order they typically encountered it. For readability in this appendix, we organize items under construct labels (e.g., Trust-Based Expectation, Satisfaction). Participants saw the survey item text itself, but not these construct labels.

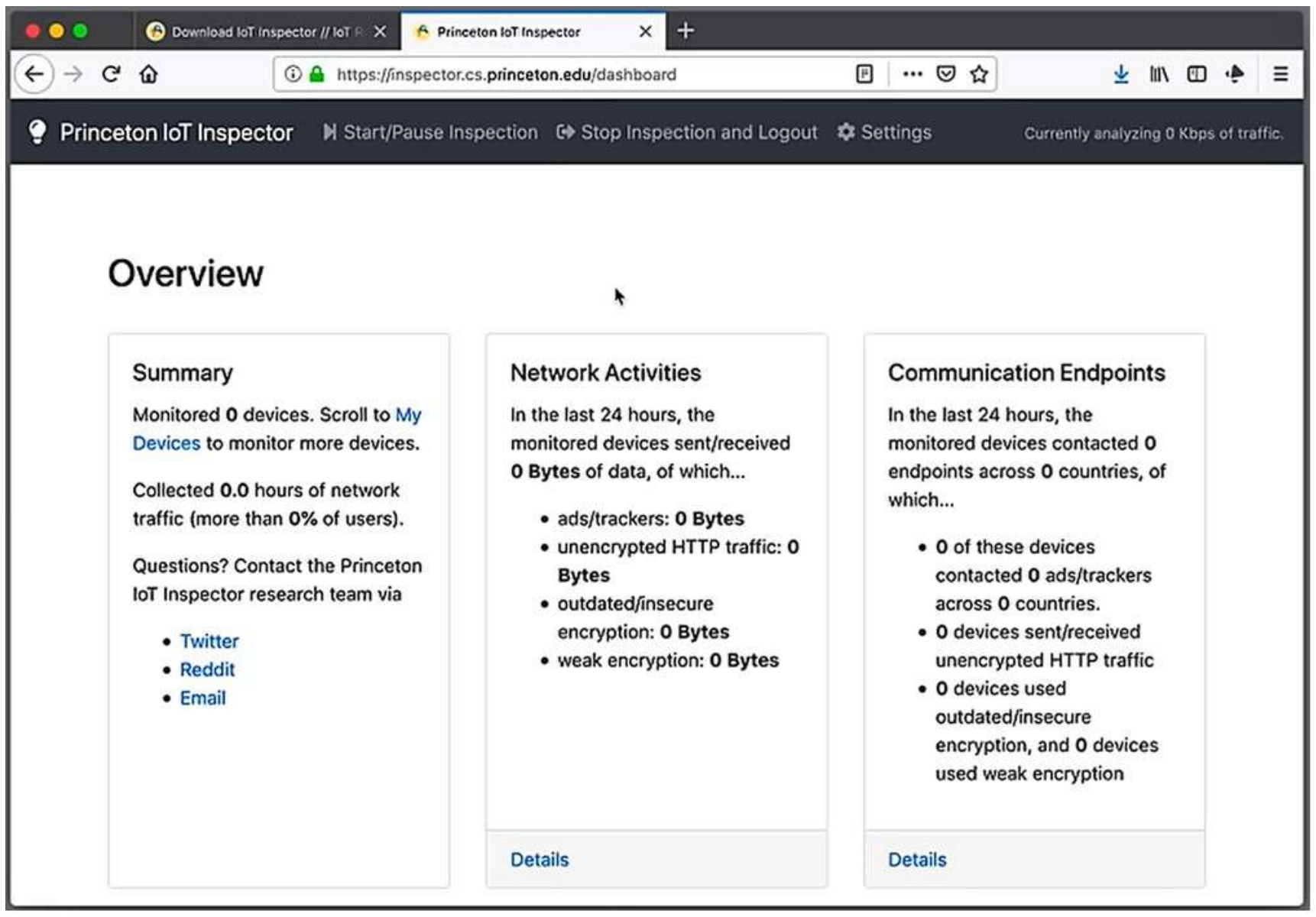

Fig. 9. A screenshot of the IoT Inspector. Users install the software and access the user interface through their browser.

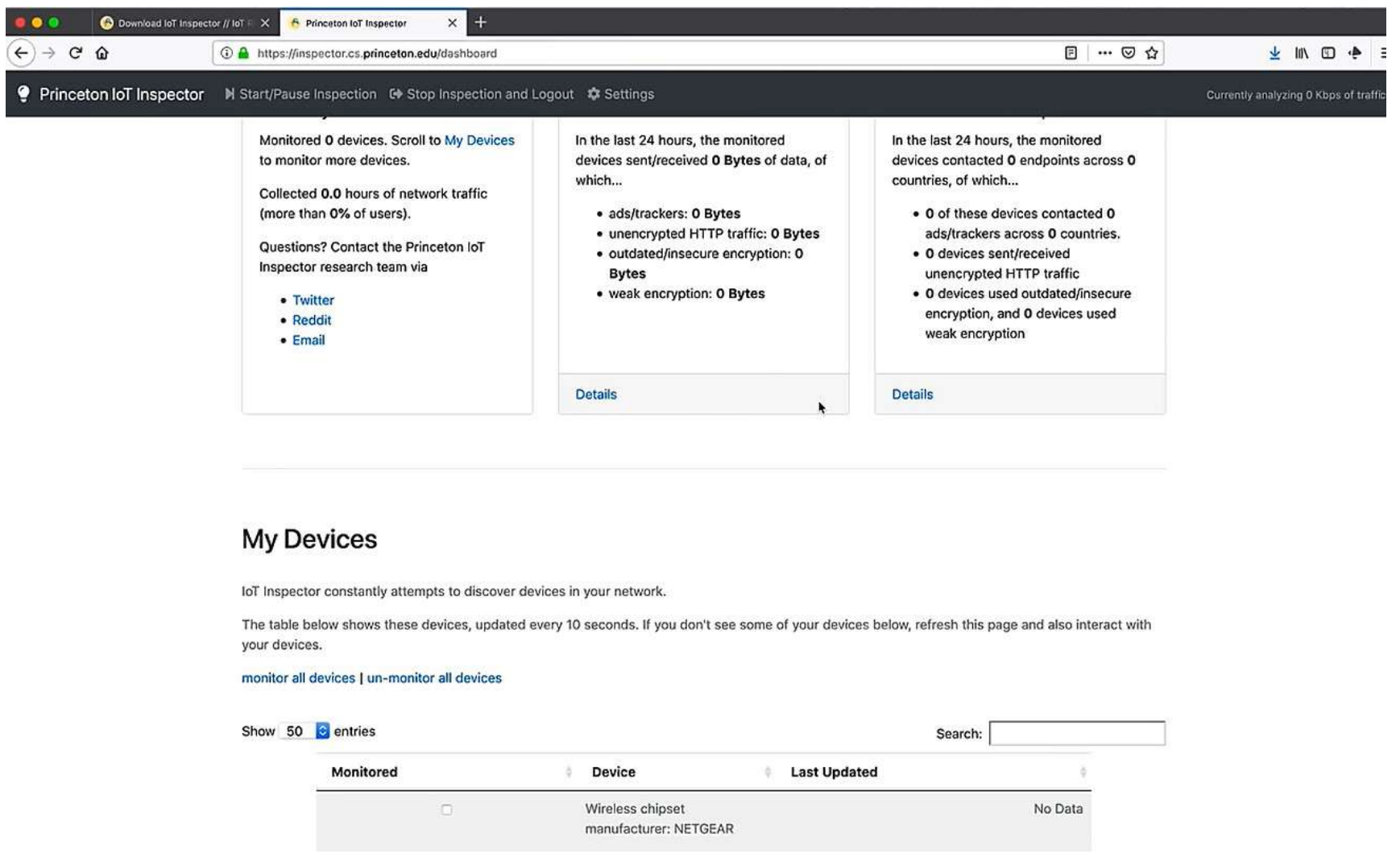

Fig. 10. A screenshot of the IoT Inspector. Users install the software and access the user interface through their browser.

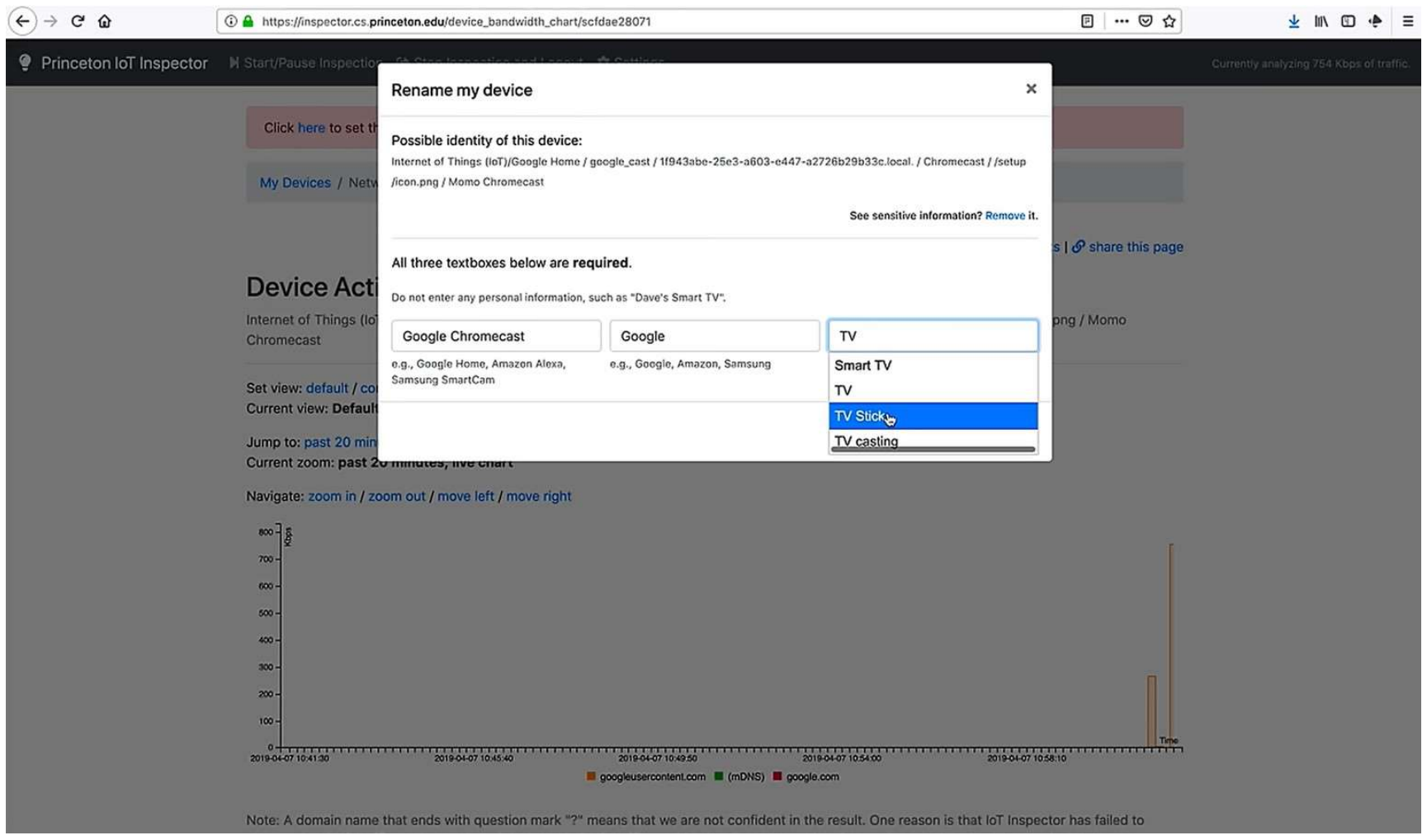

Fig. 11. A screenshot of the IoT Inspector. Users install the software and access the user interface through their browser.

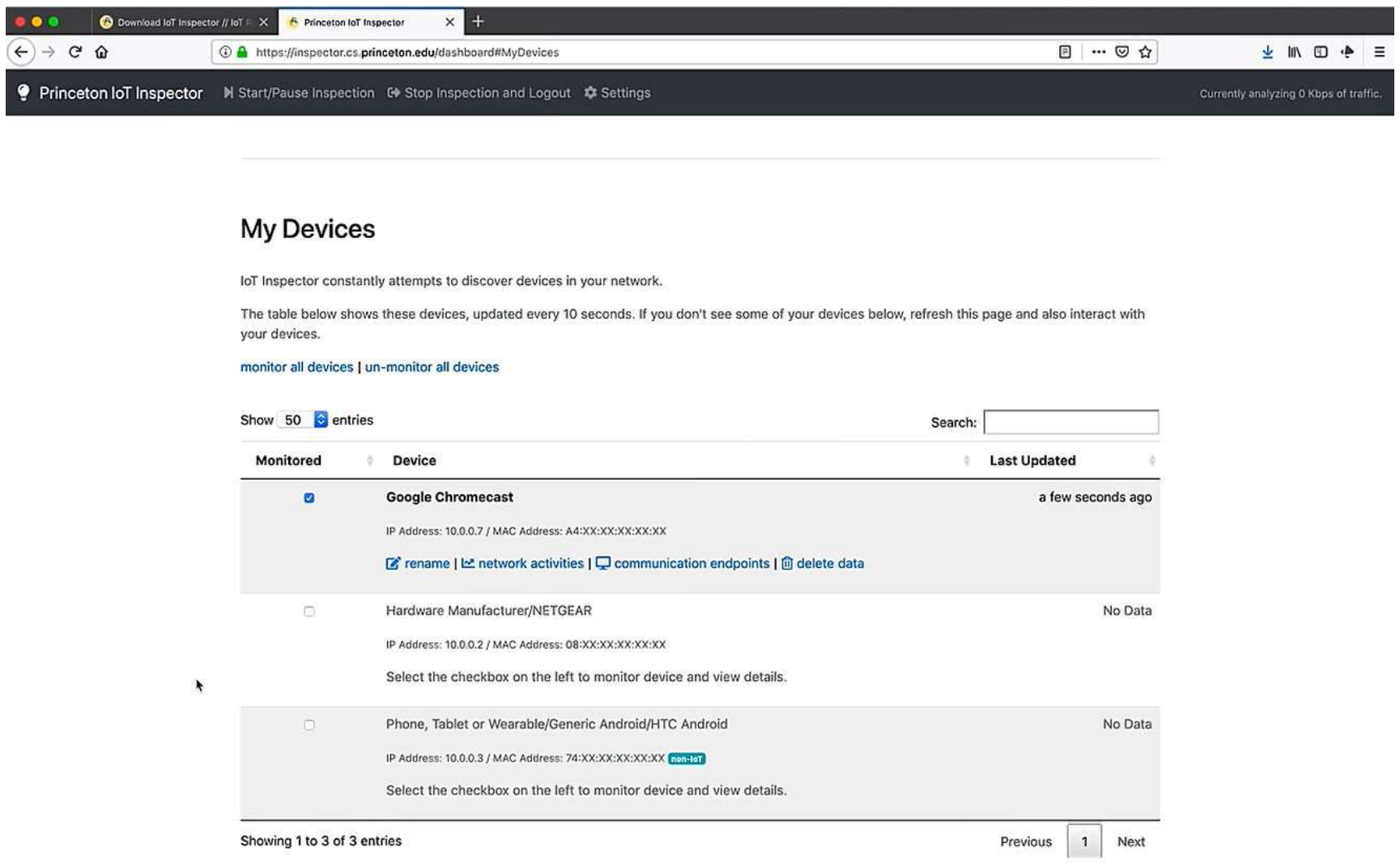

Fig. 12. A screenshot of the IoT Inspector. Users install the software and access the user interface through their browser.

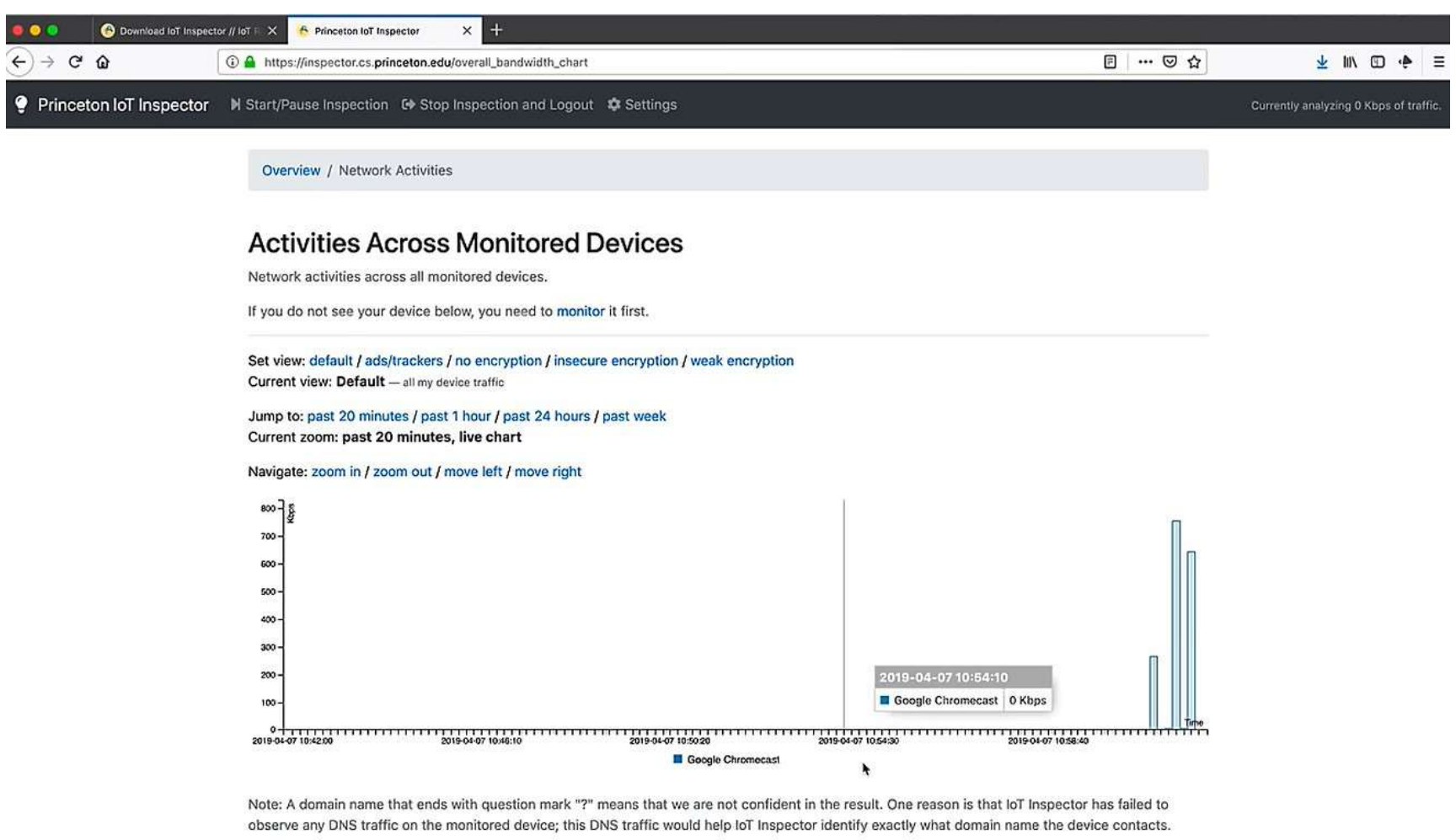

Fig. 13. A screenshot of the IoT Inspector. Users install the software and access the user interface through their browser.

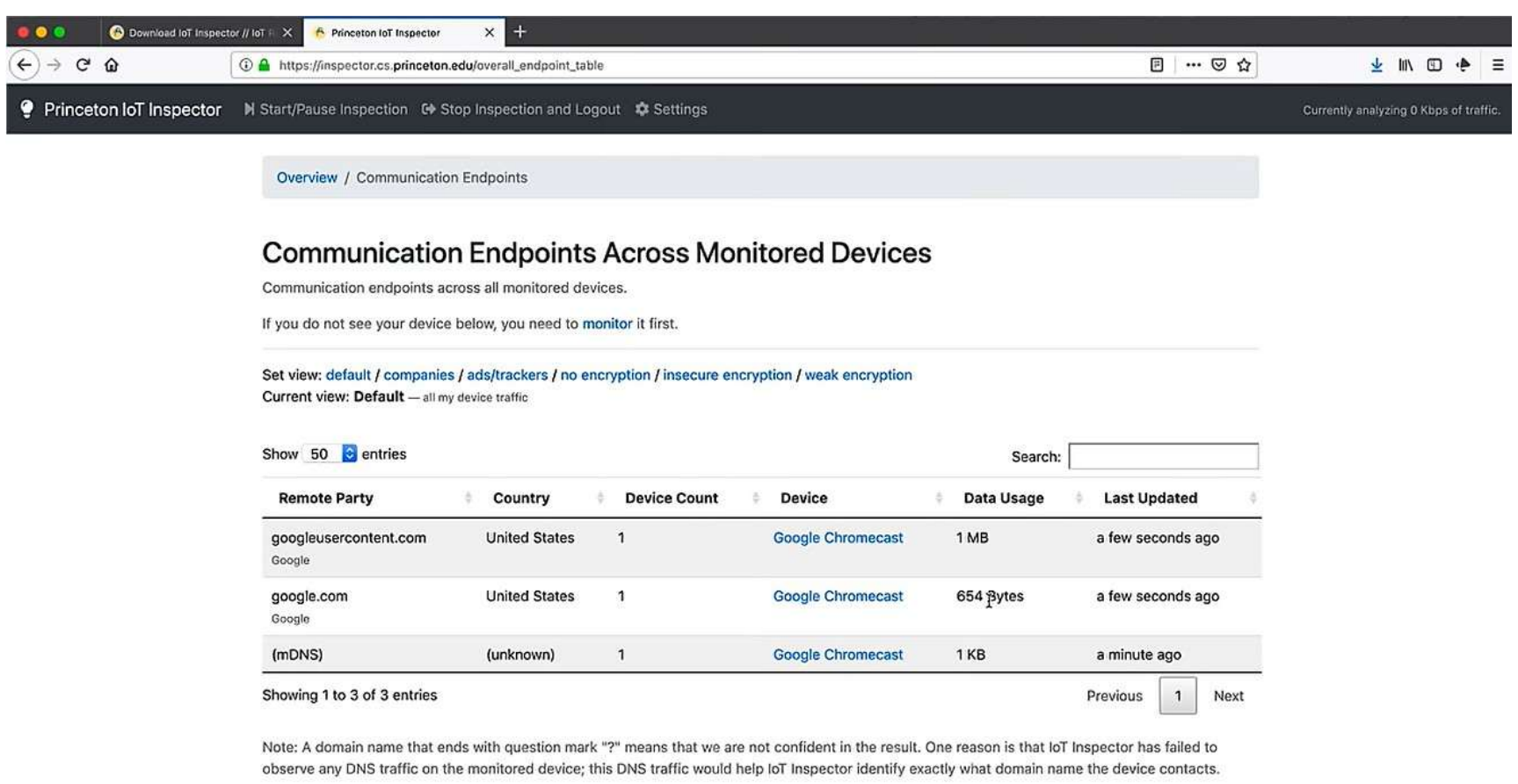

Fig. 14. A screenshot of the IoT Inspector. Users install the software and access the user interface through their browser.

### *A.2.1 Pre-Survey.*

*Informed consent.* Upon installing IoT Inspector, users saw an informed consent form asking whether they wished to participate in the study. Users who declined could continue using IoT Inspector without answering survey questions or donating network data; for these users, the tool operated locally without transmitting data to the researchers. The consent form stated that data donation included sharing the IoT Inspector report, comprising information about participants' smart devices, the frequency with which those devices send out data, and the domains with which the devices communicate.

*Trust-Based Expectation.*

- My Smart Devices keep my best interests when handling my personal data.

*Privacy Concerns.*

- All things considered, the Internet would cause serious privacy problems.

*Satisfaction.*

- I think my benefits gained from the use of my Smart Devices can offset the risks of my information disclosure.

*Intentions to Continue Use.*

- I intend to continue using my Smart Devices rather than discontinue its use.

*A.2.2 Interaction with IoT Inspector.* Participants installed IoT Inspector and browsed the report generated from their own devices' network traffic. The report included the frequency with which each smart device communicated data, the destination domain names (e.g., google.com), and the destination geographic locations (e.g., USA). Participants were prompted to complete the post-survey after at least five minutes of interaction with the report.

*A.2.3 Post-Survey.*

*Trust Confirmation.*

- Overall, most of my expectations regarding how my Smart Devices handle my personal data were confirmed. Items were coded such that higher scores indicate stronger perceived confirmation of participants' prior trust-based expectations.

*Repeated measures.* Privacy Concern, Satisfaction, Intentions to Continue Use were re-administered after the interaction with IoT inspector, using the same item wordings and scale anchors listed in the pre-survey above.

*A.2.4 Behavioral Measure: Wanting to Block.* Willingness to block was not measured with a survey item. It was captured through an interface control within IoT Inspector: a checkbox labeled "Want to Block?" displayed next to each data-collection instance for each smart device, allowing participants to mark communications they wished to prevent from recurring. This control is visible in the lower-right portion of Figure 2.

### A.3 Study 2 Survey Instrument

Items are presented below in the order in which participants encountered them. All Likert items used a fully labeled 7-point scale (1 = Strongly disagree, 2 = Disagree, 3 = Somewhat disagree, 4 = Neither agree nor disagree, 5 = Somewhat agree, 6 = Agree, 7 = Strongly agree). Throughout the survey, references to the participant's device were piped in dynamically based on the device they selected (shown below as *[device]*) and the brand they entered (shown as *[brand]*).

*A.3.1 Pre-Survey.*

*Informed consent.* Participants reviewed an informed consent form and provided explicit electronic consent before proceeding. Only participants who consented could continue.

*Device selection. Please select one of the following devices that you currently use. The rest of the survey will focus on your experiences with that device.*

- Smart Speaker
- Smart TV
- Smart Camera

*Brand. What is the brand name for your [smart device] (if you have several, please enter the one you use the most). Please don't provide any identifying information.* (open text response)

*Technology Self-Efficacy. Please specify if you agree or disagree with the following statements:*

(1) I would feel comfortable using my [smart device] on my own.
(2) If I want to, I can use my [smart device] on my own easily.

(3) I would be able to use my [smart device] even if there is no one around to show me how to use it.

*Satisfaction (Perceived Value of Information Disclosure). Please specify if you agree or disagree with the following statements:*

(1) I think my benefits gained from the use of my [smart device] can offset the risks of my information disclosure.
(2) The value I gain from use of my [smart device] is worth the information I give away.
(3) I think the benefits gained from using my [smart device] will be more than the risks of my information disclosures.

*Trust-Based Expectation.* Presented with the shared stem *"I expect my [smart device] to ..."*:

(1) ...put my interests first when handling my personal data.
(2) ...keep my interests in mind when handling my personal data.
(3) ...understand my needs and preferences when handling my personal data.

*Intentions to Continue Use. Please specify if you agree or disagree with the following statements:*

(1) I intend to continue using my [smart device] rather than discontinue its use.
(2) My intentions are to continue using my [smart device] than use any alternative means.
(3) If I could, I would like to continue my use of [smart device].

*Collection Concern. Please specify if you agree or disagree with the following statements:*

(1) It usually bothers me when online companies ask me for personal information.
(2) When online companies ask me for personal information, I sometimes think twice before providing it.
(3) It bothers me to give personal information to so many online companies.
(4) I'm concerned that online companies are collecting too much personal information about me.

*Awareness. Please specify if you agree or disagree with the following statements:*

(1) Companies seeking information online should disclose the way the data are collected, processed, and used.
(2) *Please select "Disagree"*[4]
(3) A good consumer online privacy policy should have a clear and conspicuous disclosure.
(4) It is very important to me that I am aware and knowledgeable about how my personal information will be used.

*Unauthorized Use. Please specify if you agree or disagree with the following statements:*

(1) Online companies should not use personal information for any purpose unless it has been authorized by the individuals who provided information.
(2) When people give personal information to an online company for some reason, the online company should never use the information for any other reason.
(3) Online companies should never sell the personal information in their computer databases to other companies.
(4) Online companies should never share personal information with other companies unless it has been authorized by the individuals who provided the information.

*A.3.2 Stimulus Presentation.* Participants were first shown the message: *"Now we will analyze your network data to tell you about the data practices of your [smart device],"* followed by an animated progress bar displaying *"Analyzing your network data..."*

Participants were then randomly assigned to view either the LowAd or the HighAd report. Both reports were introduced with identical text: *"Below, you can see how much data your [device] communicated with [brand] and with ADVERTISEMENT-related domains: All entities contacted by this device over the last 10 minutes."* Each report displayed traffic volume for the manufacturer server and for advertisement domains, followed by the prompt *"If you want to block the Advertisement Domain click the checkbox below"* and a single response option, *"Block communications with Advertisement Domains."*

[4] This item was an attention check embedded within the Awareness block and was not included in the Awareness construct.

- LowAd condition: [brand] Server – 842 KB uploaded, 392 KB downloaded; Advertisement Domains – 90 KB uploaded, 110 KB downloaded.
- HighAd condition: [brand] Server – 842 KB uploaded, 392 KB downloaded; Advertisement Domains – 1.2 MB uploaded, 900 KB downloaded.

The full reports as participants saw them:

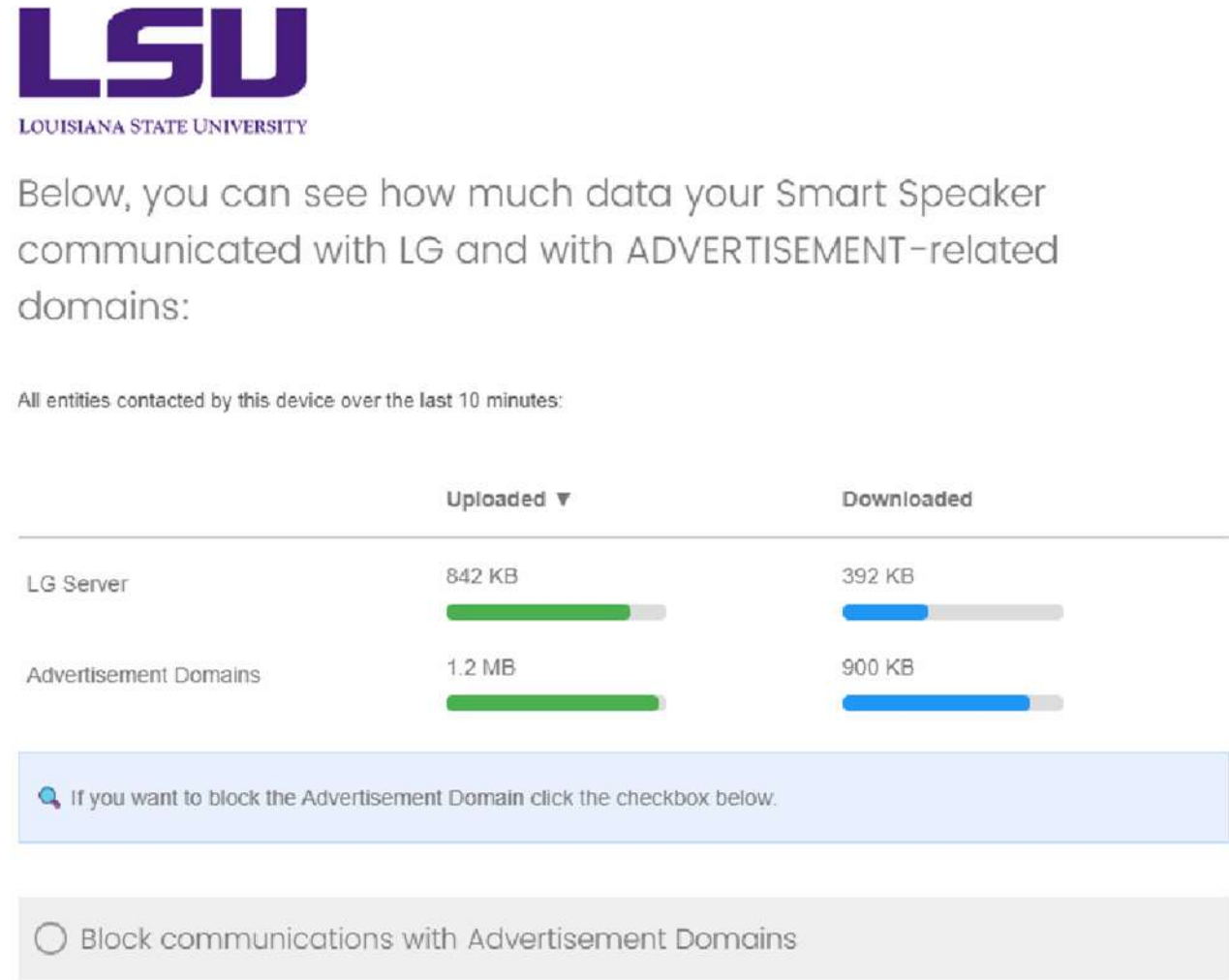

Fig. 15. Simulated network activity reports comparing the HighAd conditions. Advertisement traffic volume is displayed relative to the manufacturer baseline (e.g., LG or Samsung servers), with a radio button as a choice option for blocking communications.

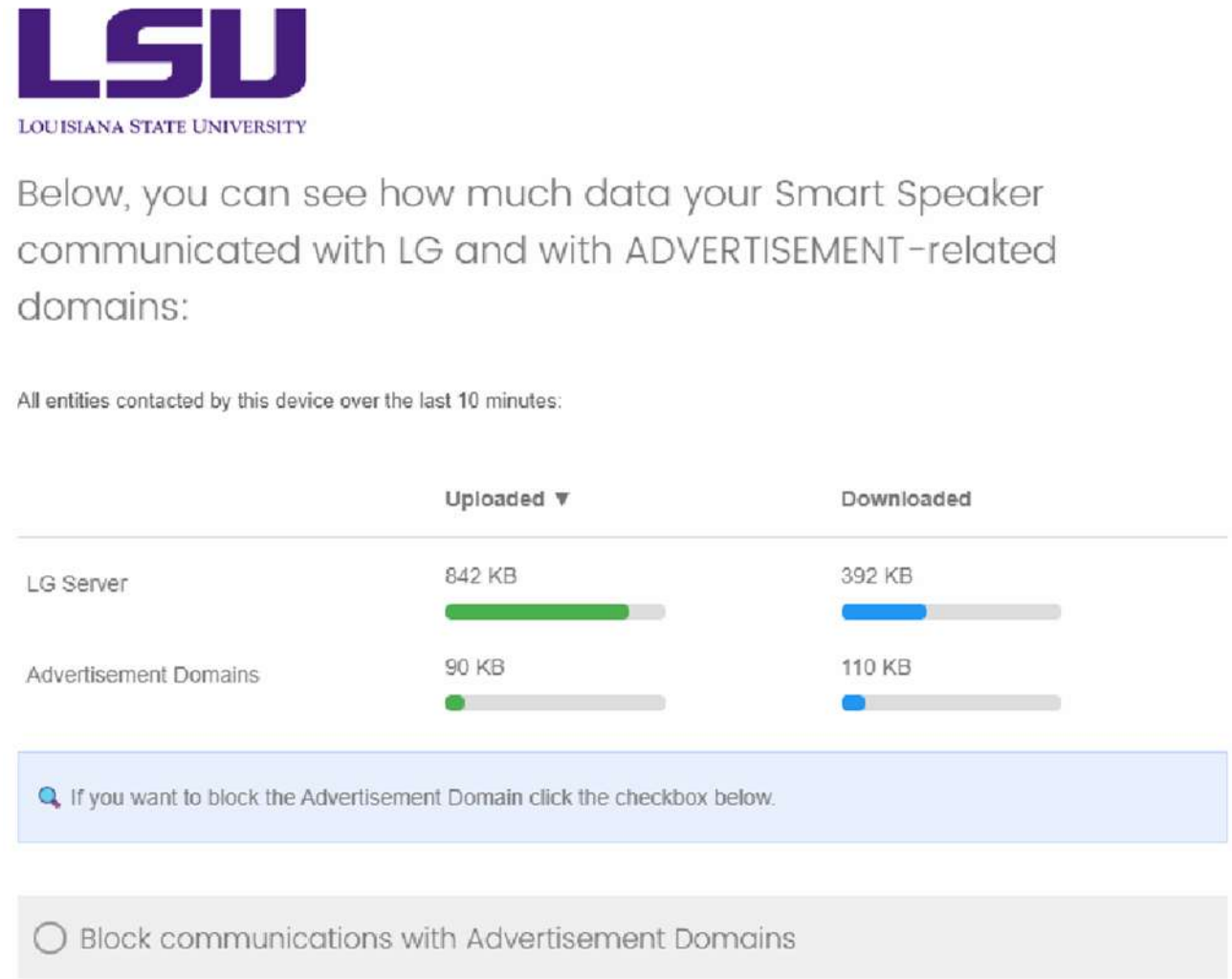

Fig. 16. Simulated network activity reports comparing the LowAd conditions. Advertisement traffic volume is displayed relative to the manufacturer baseline (e.g., LG or Samsung servers), with a radio button as a choice option for blocking communications.

*A.3.3 Post-Survey.* Participants were introduced to the post-survey with the message *"Thanks for responding to our questions. Here are a few more questions."*

*Trust Confirmation.* (post-survey only) *Please specify if you agree or disagree with the following statements:*

(1) The way my [smart device] handled my personal data was better than what I expected.
(2) Overall, most of my expectations regarding how my [smart device] handle my personal data were confirmed.[5](3) My experience confirmed that my [smart device] put my interest first when handling my personal data.
(4) My experience confirmed that my [smart device] keep my interests in mind when handling my personal data.
(5) My experience confirmed that my [smart device] want to understand my needs and preferences when handling my personal data.

*Repeated measures.* Self-Efficacy, Satisfaction, Intentions to Continue Use, Collection Concern, Awareness, and Unauthorized Use were re-administered after the manipulation, in that order, using the same item wordings and scale anchors listed in the pre-survey above. An attention-check item (*"Please select 'Strongly Agree' "*) was embedded within the Collection Concern block and was not included in the construct.

*Open-ended and usage questions.*

(1) *Were the data practices of your [smart device] consistent with your expectations? Please explain. Please don't provide any identifying information.* (open text response)
(2) *How frequently do you use your [smart device]?* (About once a week / Several times a week / At least once a day / Several times a day / Almost constantly)
(3) *What do you use your [smart device] for? Please provide examples. Please don't provide any identifying information.* (open text response)
(4) *What did you think about the report we showed you on your network traffic, and did you want to see any other information in the report? Please don't provide any identifying information.* (open text response) The thematic analysis reported in Section 5.2.4 analyzed responses to the first of these questions.

*Additional exploratory items.* Two further items were administered but were not analyzed in this paper:

(1) The report that I saw about data practices of my [smart device] improved my performance in managing my home.
(2) The report that I saw about the data practices of my [smart device] improved my performance in managing my Privacy.

*Demographics.*

- Education. *What is the highest level of school you have completed or the highest degree you have received?* (High school incomplete or less; High school graduate or GED (includes technical/vocational training that doesn't count towards college credit); Some college (some community college, associate's degree); Four year college degree/bachelor's degree; Some postgraduate or professional schooling, no postgraduate degree; Postgraduate or professional degree, including master's, doctorate, medical or law degree)
- Race. *Which of the following describes your race? You can select as many as apply.* (White; Black or AfricanAmerican; American Indian or Alaska Native; Asian or Asian-American; Native Hawaiian/Other Pacific Islanders; Some other race)

[5] Removed from the final measurement model due to a low standardized factor loading (see Section 5.2.1).

- Income. *What is your total household income?* (Less than $10,000; $10,000 to less than $20,000; $20,000 to less than $30,000; $30,000 to less than $40,000; $40,000 to less than $50,000; $50,000 to less than $75,000; $75,000 to less than $100,000; $100,000 to less than $150,000; $150,000 or more)
- Age. *Please enter your age. Please don't provide any identifying information.* (open text response)
- Gender. *What is your gender identity?* (Woman; Man; Non-binary; Prefer not to say)

*Debriefing and data-retention consent.* Participants were told: *"We want to let you know that some parts of the study involved simulated scenarios. Specifically, the network diagram you saw about your [device] was fictional and not based on actual data from your device. This was done to better understand how people perceive privacy based on different types of information presentation. If you have any concerns or would prefer that we delete your responses, you may request that now—your decision will not affect your compensation in any way."* Participants then selected either *"I want my responses being used for research"* or *"I DO NOT want my responses being used for research."*

## A.4 Qualitative Analysis

We have restructured the qualitative results section and added more detail about the sub-themes, including how participants arrived at their judgments of consistency (Section 5.2.5, Table 7).

*A.4.1 Coding procedure for Study 2.* For the qualitative analysis, we followed Braun and Clarke's six-phase thematic analysis approach [13]. Two researchers first familiarized themselves with the open-ended responses by reading through them independently. They then generated initial codes across the dataset, capturing recurring interpretations participants gave for whether the reported data practices matched their expectations. Codes were collated into candidate themes, which were reviewed against the coded extracts and the full dataset, refined, and defined. Through iterative discussion, the researchers consolidated the codes into three primary themes: Data Practices Consistent with Expectation, Distrust Toward Data Practices, and Others, two of which contained interpretive sub-themes reflecting how participants arrived at their judgments. Differences in coding and interpretation were discussed and refined through consensus.

The final codebook resulting from this process is presented in Table 9, including the themes and subthemes, their coding rules, and illustrative indicators. Top-level themes were mutually exclusive, whereas subthemes could co-occur within a response.

Table 9. Qualitative codebook for Study 2.

| Code | Coding rule | Illustrative indicator |
|---|---|---|
| Data Practices Consistent with Expectation | Apply when the participant's overall judgment is that the depicted data practices broadly matched their prior expectations. A response could also receive one or more of the interpretive subtheme codes below. | "Yes"; "matched my expectations." |
| Normalization or pragmatic acceptance | Apply when the participant describes the depicted data sharing as routine, normal, or unavoidable for smart devices. Consistency reflects resignation or pragmatic acceptance rather than approval. | "All smart devices are using our data." |
| Prior-knowledge confirmation | Apply when the participant identifies specific data practices they already knew about or expected, such as collection of viewing history, app usage, voice commands, or recordings. | "I was aware it would collect viewing history." |
| Favorable deviation from expectations | Apply when the participant reports that the depicted communication was lower, less extensive, or less concerning than expected and expresses relief or reassurance. | “A bit lower than my expectations.” |
| Overall consistency with partial surprise | Apply when the participant gives an overall judgment of consistency but also identifies a particular communication or practice that was surprising, concerning, or deserving of greater transparency or control. | “Somewhat consistent, but I was surprised.” |
| Distrust Toward Data Practices | Apply when the participant indicates that the depicted practices did not match their expectations and expresses suspicion, discomfort, or distrust toward the device’s data handling. A response could also receive one or more of the subtheme codes below. | “Not consistent with my expectations.” |
| Volume exceeded expectations | Apply when the participant’s distrust focuses on the amount, frequency, or intensity of communication shown in the report, including communication with advertising-related domains. | “So much data in the last 10 minutes.” |

*Continued on next page*

*Table 9 continued*

| Code | Coding rule | Illustrative indicator |
|---|---|---|
| Protection or consent violated | Apply when the participant says that the device should have protected their data, requested permission, or avoided using or sharing the data without consent. | “I expected to be asked for permission.” |
| Others | Apply when the response does not provide a discernible judgment about the depicted data practices, such as when it is neutral, unrelated to the report, or indicates no prior expectation. Brief responses that clearly indicated consistency or inconsistency were assigned to the corresponding top-level theme but were not assigned an interpretive subtheme. | “I didn’t really have any expectations.” |